%% file: main.tex
\documentclass[aps,prx,showpacs,twocolumn]{revtex4-2} 
\usepackage{amsmath,amssymb}
\usepackage{bm}
\usepackage{tipa}
\usepackage{upgreek}

\usepackage{comment}
\usepackage{mathrsfs}
\usepackage{float}
\usepackage{mathrsfs}
\usepackage{graphicx}
\usepackage{braket}
\usepackage{mathbbol}
\usepackage{booktabs}
\usepackage{amssymb}
\usepackage{enumitem}
\usepackage[normalem]{ulem}
\usepackage{arydshln}
\usepackage{color}
\usepackage[table,xcdraw]{xcolor}
\usepackage[colorlinks,bookmarks=true,citecolor=blue,linkcolor=red,urlcolor=blue]{hyperref}
\usepackage{cleveref}

\makeatletter
\let\old@makecaption=\@makecaption
\usepackage{subcaption}
\let\@makecaption=\old@makecaption
\makeatother
\usepackage{relsize}

\newcommand{\chern}{\mathcal{C}}
\newcommand{\SU}{\text{SU}}
\newcommand{\U}{\text{U}}

\begin{document}

\title{Phases of itinerant anyons in Laughlin's quantum Hall states on a lattice}

\author{Tevž Lotrič and Steven H. Simon}
\affiliation{Rudolf Peierls Centre for Theoretical Physics, Parks Road, Oxford, OX1 3PU, United Kingdom}
%\date{\today}

\begin{abstract}
    We study phases of itinerant anyons when hole-doping Laughlin-like states in fractional Chern insulators (FCIs). In light of the recent observation of time-reversal-broken superconductivity near FCIs in van der Waals materials,  a theoretical understanding of doped fractional quantum Hall states on a lattice has been developed by Shi and Senthil [Phys. Rev. X 15, 031069], reviving old ideas about ``anyon superconductivity''.  We test these ideas analytically within an effective parton mean-field theory and numerically with variational Monte Carlo, pointing out that the predicted state depends on whether the Laughlin order at $\nu=1/m$ is described by a $\U(1)$, or an $\SU(m)$ Chern-Simons field, the latter implying a symmetry between the $m$ parton species. Our results demonstrate that the interplay between band Berry curvature and effective anyon dispersion has crucial implications for which anyonic phase is realized. In the experimentally relevant scenario of hole-doping the $\nu=1/3$ fermionic FCI, our results uncover a mechanism for the formation of an anyon superconducting state of half-integer central charge in the case when the energetically cheapest excitations are the fundamental $1/3$ charge anyons, bypassing the need for these anyons to pair into charge-$2/3$ composites, which has generally been assumed in similar anyon superconductivity constructions.
\end{abstract}
\maketitle

\section{Introduction} 
Fractional Chern insulators (FCIs) are analogues of fractional quantum Hall (FQH) states in the presence of a periodic lattice potential~\cite{doi:10.1142/S021797921330017X,parameswaran_review_2013,LIU2024515}. The basic observation leading to one class of FCIs is that most of the properties of FQH states, such as topological order and anyonic excitations, should not change when a magnetic field in a uniform background is replaced by its lattice version, a (nearly) flat topological band with Chern number $\chern=1$.  This $\chern=1$ condition is in fact too strict to account for all known FCI's \cite{M_ller_2015,Lauchli2011,SimonHarper,Lin_2025,lupreprint}, but it is the one case most relevant to Van der Waals materials (MoTe$_2$ and rhombohedral pentalayer graphene \cite{xu_signatures_2025,han_signatures_2025,xu_observation_2023,park_observation_2023,lu_fractional_2024}), where for the first time FQH-like states have been detected in zero applied magnetic field. Upon shifting the chemical potential away from the FCI plateau, the systems display superconductivity, which is surprising in light of the strong sense of time-reversal symmetry breaking required to stabilize an FCI. Recently, a physical explanation for this superconducting state has emerged \cite{senthil_doping_2024,senthil_doping_2025,sahay_superconductivity_2024,kim_variational_2025,soejima_anyon_2024,song_doping_2021,wen_topological_2025,schleith_anyon_2025}, reviving old ideas about anyon superconductivity \cite{Chen:1989xs,WEN1990135,PhysRevB.39.9679}. The crucial insight is that unlike in the quantum Hall problem, where the magnetic field quenches the motion of anyonic excitations, the breaking of continuous translation symmetry by the lattice in an FCI allows the anyons to develop a dispersive kinetic energy depending on an appropriately defined Bloch momentum \cite{yan_anyon_2025}. This makes the anyons itinerant and allows them to form many-body phases of their own, stacked on top of the FCI's topological order. Itinerant anyons are driven to minimize their kinetic energy $\epsilon\propto \xi^{-2}$ by delocalizing over a large area of linear size $\xi$. But at a finite anyon density $\delta$, an anyon delocalizing over $\xi\gtrsim 1/\sqrt\delta$ starts to feel the exchange phases from other nearby anyons, introducing frustration to the kinetic energy. It is this competition between kinetic energy minimization and the resulting penalty from the statistical interaction that determines the physics of a gas of anyons. \textit{Anyon superconductivity} is a mechanism in which the anyons form clusters which together feel no statistical phase, allowing them to delocalize freely and minimize the kinetic energy. These clusters are the carriers of the dissipationless supercurrent.

 Exploring the physics of itinerant anyons opens the possibility of new physical behaviours and processes beyond those emerging from bosons or fermions. At the same time, studying these systems analytically or numerically presents significant challenges, as the constituent particles, the anyons, can only emerge from an already highly-correlated system and themselves have no free description. Refs.~\cite{senthil_doping_2024,senthil_doping_2025} have proposed a description of this mechanism and the resulting states within effective Chern-Simons field theory, which while clarifying many aspects of anyon superconductivity, is generally highly indifferent to the details of microscopic energetics and thus cannot make precise statements about what states are energetically favoured over others.   On the opposite end, numerical Density-Matrix Renormalization Group (DMRG) calculations have found superconducting behaviour for system parameters close to those of an FCI, but the identification of the result as an anyon superconductor has been difficult, in part due to a mismatch of the numerically found pairing symmetry and the chiral central charge predicted in theory \cite{wang_chiral_2025}. Bridging the gap between the two is one significant contribution of this work, as we aim to more directly test the field-theoretical predictions.  Furthermore, we 
propose new states of doped FCIs which might agree better with other numerical methods. 
Our work introduces analytical methods for understanding doped FCI states and establishes a special type of variational Monte Carlo (VMC), which is an extension of our previous work in Ref.~\cite{lotric_paired_2025}, as a useful tool for numerically studying doped FCI states, extending the scope of VMC methods and bringing the analytical and numerical results for the doped FCI problem closer together.

In Section~\ref{sect:parton_mf_theory}, we first review the parton construction-based theory of Ref.~\cite{senthil_doping_2024,senthil_doping_2025}, pointing out that one of the crucial assumptions of those works might be violated due to the large $\SU(m)$ gauge invariance present in the parton construction of Laughlin's state and its generalizations in some special lattice models. We then present a large class of new states that may emerge upon doping this  state and use a mean-field approach to establish a picture of how microscopic effects such as particle interactions and the interplay between the geometry and dispersion in the bands are crucial for determining the fate of the doped system.

In Sect.~\ref{sect:vmc}, we test our theory with variational Monte Carlo which finds the lowest-energy wavefunction among a wide range of non-interacting parton trial states.  This type of Monte Carlo has recently been shown to be extremely accurate compared to exact diagonalization for certain FCIs \cite{lotric_paired_2025}.  Our results largely confirm the expectations set by the mean-field picture. We find cases where our numerics confirm the predictions of \cite{senthil_doping_2024,senthil_doping_2025}, while in other cases we discover new behaviours of anyons upon doping.  We discuss some of these phases in more detail in Sect.~\ref{sect:undistr} and finally in Sect.~\ref{sect:discussion} we conclude by placing our results in the context of other work and experimental results, focusing on one new prediction of our work, a mechanism for obtaining an anyon superconductor without the need for the $Q=\frac13$ anyons to first pair into larger $Q=\frac23$ objects.
This proposed superconducting state has a half-integer chiral central charge, which is, in the weak pairing limit , consistent with an odd integer pairing angular momentum and \textit{triplet} pairing \cite{46_Read_Green}, which is the expected pairing symmetry for spin-polarized electrons.

\section{Parton theory of doped FQH states} \label{sect:parton_mf_theory}

 We first review the parton construction, which underlines both the existing field-theoretical predictions, and also the variational forms used in the present work. We note that partons may be defined either on a lattice or in a continuum system. Most of the theory we discuss in the following sections is most easily phrased in the continuum Landau level, although parts are more transparent on a lattice. We note that a wide class of lattice tight-binding models can be mapped onto such a continuum picture, meaning that any conclusions we arrive at in the continuum also hold for those lattice models. For concreteness, the Kapit-Mueller model \cite{kapit_exact_2010} allows one to convert between the two pictures. We turn to a generalization of this model for the numerical calculation in Sect.~\ref{sect:vmc}.

Use of the parton construction in the quantum Hall problem has a long history~\cite{PhysRevB.40.8079,wen_book,11_Wen_1999,05_barkMcGreevyTheory,25_McGreevy_wavefunction_PhysRevB.85.125105,Balram_2024,PhysRevX.12.021008,Barkeshli_2012,15_Senthil_theory} and in this case it may be motivated by noting that the $\nu=1/m$ Laughlin wavefunction (describing bosons for $m$ even and fermions for $m$ odd) may be written as a product of $m$ Slater determinants,
\begin{equation} \label{eq:laughlin_as_partons}
    \prod_{i<j}(z_i-z_j)^m e^{-\frac{B}{4}\sum_i |z_i|^2}=\left[ \prod_{i<j}(z_i-z_j) e^{-\frac{B}{4m}\sum_i |z_i|^2} \right]^m
\end{equation}
suggesting that we may write the microscopic boson/fermion annihilation operator as $\psi_{b/f}=\prod_{p=1}^m f_p$ where $f_p$ are fermionic parton annihilation operators which may be thought of as being roughly free at some mean-field level. If we are dealing with microscopic bosons, the parton decomposition enforces a hard-core constraint $\psi_b^2=0$. In Eq.~\ref{eq:laughlin_as_partons}, each of the $m$ partons $f_p$ fills a lowest Landau Level to give Laughlin's fractional quantum Hall state, but more generally if all $m$ partons independently fill Chern $\chern=1$ bands, the resulting state has the same topological order and would be refered to as a fractional Chern insulator.

Given a microscopic Hamiltonian $H[\psi,\psi^\dagger]$, the challenge is to first find what type of parton state leads (close) to the ground state (if partons are even a good description), and then to infer what phase of matter that parton state corresponds to. For the first step, we start by making the substitution $\psi\rightarrow\prod_{p=1}^mf_p$ in $H$. This only makes the problem more difficult, but the next crucial step is to make a Hartree-type approximation, assuming that partons are in a Slater determinant state and that the only non-zero partonic bilinears are $\langle f_p^\dagger f_q\rangle$ with $p=q$. One should be able to use this approximation to find some self-consistent mean-field parton ground state and a corresponding quadratic Hartree Hamiltonian $H_\text{MF}^{(p)}=\int_{\mathbf{x},\mathbf{y}}t_p(\mathbf x,\mathbf y)f_p^\dagger(\mathbf x)f_p(\mathbf y)$ for each parton $p$. This is not guaranteed to be a good approximation to the ground state, but it clearly describes a different set of mean-field saddle points than for example a direct Hartree approximation on $\psi$ and this construction turns out to be very useful when studying fractionalization. While the parton mean-field is not a controlled approximation, the predictions it makes may always be checked by constructing microscopic variational wavefunctions which may then be compared to results obtained from other numerical techniques. In Ref.~\cite{lotric_paired_2025}, we studied the $\nu=\frac12$ undoped bosonic problem, where we found the parton description to be very effective. We found that this parton \textit{Ansatz} may evolve to a superfluid phase if parameters of $H$ are tuned and that when this happens, the parton decomposition only remains a good approximation if an anomalous term $\sim \Delta f_1f_2$ is added to $H^\text{MF}_p$. In the current work however, we choose to ignore such corrections, which we reason as follows: such anomalous terms may only be added for the $m=2$ bosonic case, as in $m>2$ cases, any anomalous parton bilinear would ruin the internal gauge invariance discussed below in Sect.~\ref{sect:parton_gauge_fields}. Even in the $m=2$ case, we expect $\Delta$ to be related to the superfluid weight \cite{lotric_paired_2025} which is in turn proportional to the dopant density $\delta$ and thus small. This means that even in the $m=2$ case, the effect of such anomalous terms should not be large as compared to what was observed in Ref.~\cite{lotric_paired_2025}.

\subsection{Internal gauge fields} \label{sect:parton_gauge_fields}

There is an immediate problem concerning the Hilbert spaces of the microscopic problem and the parton mean-field, which is most transparent when regularizing to a lattice with $N_s$ sites \cite{wen_book}. As $\psi$ describes either fermions or hard-core bosons, there are $\dim\mathcal{H}=2^{N_s}$ states. But the partons see a much larger space $\dim\mathcal{H}^\text{MF}=2^{mN_s}$, meaning that most of the parton Hilbert space does not correspond to any physical state. Denoting with $\hat n_p(\mathbf x)=f_p^\dagger(\mathbf x)f_p(\mathbf x)$ the number operator of parton $p$ on physical site $\mathbf x$, a parton state is only physical if $\hat{n}_1(\mathbf x)=\hat n_2(\mathbf x)=\ldots=\hat{n}_m(\mathbf x)$ for all $N_s$ sites $\mathbf x$. In field theory, this is enforced by coupling the partons to an $\SU(m)$ internal gauge field, where $(f_1,\ldots,f_m)^T$ transforms in the fundamental representation~\cite{wen_book}. The effective low-energy parton theory must include both the excitations of partons around their mean-field ground states, as well as fluctuations of this $\SU(m)$ gauge field which effectively ``glue'' the partons together to fulfil the condition of being in the physical subspace. The $\SU(m)$ gauge field can rotate between different parton flavours and shows up in the effective low-energy theory only when the partons all have an identical mean-field \textit{Ansatz}. In the case that parton \textit{Ansätze} are different, the field may be ``Higgsed'' with only a sub-group remaining in the low energy theory --- in our case we will observe $\SU(m)\rightarrow \U(1)^{m-1}$ and also $\SU(3)\rightarrow \U(2)$. 

Integrating over the gauge fields is itself a difficult problem, but if the ground state of $H_\text{MF}^{(p)}$ is gapped for all partons $p$, we may to lowest order replace the gauge field by a constant mean-field value, giving us an entirely free-fermion parton picture at zeroth order \cite{wen_book,11_Wen_1999}. If one of the parton states is gapless however, fluctuations in the gauge fields are important.

In a Landau level in a magnetic field $B$ at particle filling $\nu=1/m$, the mean-field corresponds to an $\SU(m)$ flux $\frac B m\text{diag}[1,1,\ldots,-(m-1)]$ if we assign external charges $(0,0,\ldots,1)$ to the partons (the full unit charge must be distributed among the partons) and assume that the parton mean-field states are \textit{identical}. Alternatively, if (unlike in the Laughlin state) the $m$ partons were \textit{different}, we would have $m-1$ different $\U(1)$ gauge fields $a_1,\ldots,a_{m-1}$ with parton $f_1$ charged under $a_1$, partons $f_p$ under $a_{p}-a_{p-1}$ for $p=2,\ldots,m-1$ and parton $f_p$ under $A-a_{m-1}$. The mean-field in this case is $\langle \nabla \wedge a_k\rangle=\frac{k}{m}\langle \nabla\wedge A \rangle$. In both cases,  each parton effectively lives in a field $B/m$, giving it a filling fraction $\nu_p=1$. Each parton fills a Landau Level and is thus gapped at the mean-field level, meaning that we may at first ignore the gauge field to yield a parton MF state $\ket\Omega=\ket{\Omega_1}\ldots\ket{\Omega_m}$ with each $\ket{\Omega_p}$ a filled lowest Landau level (LLL) at field $B/m$. To asses the accuracy of this approximation, we may construct a microscopic wavefunction out of the mean-field state $\ket\Omega$ as $\psi(\mathbf r_1,\mathbf r_2,\ldots)=\braket{\emptyset|\psi(\mathbf{r_1})\psi(\mathbf r_2)\ldots|\Omega }$. Upon replacing $\psi$ with the parton decomposition, we get
\begin{equation} \label{eq:general_parton_to_wf}
    \psi(\mathbf r_1,\mathbf r_2,\ldots) = \prod_{p=1}^m \braket{\emptyset|f_p(\mathbf r_1)f_p(\mathbf r_2)\ldots|\Omega_p}.
\end{equation}
Replacing all $\ket{\Omega_p}$ with the filled LLL and comparing to Eq.~\ref{eq:laughlin_as_partons} shows that our mean field together with Eq.~\ref{eq:general_parton_to_wf} indeed recovers the Laughlin state exactly. But this approach is more general --- given any microscopic Hamiltonian and parton decomposition, we may construct a variational approximation to the ground state via Eq.~\ref{eq:general_parton_to_wf}. We are therefore considering general many-body mean-field wavefunctions of the form
\begin{equation} \label{eq:partonform}
    \psi(\mathbf r_1,\mathbf r_2,\ldots) = \prod_{p=1}^m  \Omega_p(\mathbf r_1,\mathbf r_2,\ldots ) 
\end{equation}
where $\Omega_p$ is a noninteracting (single Slater determinant) wavefunction for the $p^{th}$ species of parton.  In general, no symmetry between the partons is enforced in our calculations. But in the cases in Sect.~\ref{sect:vmc} where we claim a state has $\SU(2),~\U(2)$ or $\SU(3)$ gauge invariance, we have checked that a parton \textit{Ansatz} with the appropriate symmetry re-produces the same energy (to enforce $\U(2)$ with $m=3$ for example, we need $\Omega_1=\Omega_2\neq\Omega_3$).  While parton mean-field theory, which we turn to in Sect.~\ref{sub:senthil_cstheory},~\ref{sub:mfdope}, can be used to propose different states, the derived wavefunctions of the form Eq.~\ref{eq:partonform} are ultimately the best available test of which state is closest to the true ground state.

\subsection{Framework for doped states} \label{sub:senthil_cstheory}
Our interest in this work lies in what happens when the filling is tuned slightly away from one of the Laughlin plateaus. In particular, we focus on the hole-doped case of filling $\nu=\frac1m-\delta$ with $0<\delta\ll1$. A parton-based framework for understanding these states was established in Ref.~\cite{senthil_doping_2024}, and we review part of that argument here. 

Consider bosons at $\nu=\frac12-\delta$ in a Landau level with a weak periodically modulated background potential $V(\mathbf r)$ such that there is one flux per unit cell of the modulation. We assume that the undoped state at $\nu=\frac12$ is well-captured by a parton construction $\psi_b=f_1f_2$ where the two partons are inequivalent (the case of equal partons will be treated later, in Sect.~\ref{sub:mfdope}). The inequivalence of the partons reduces the gauge group from $\SU(2)$ to $\U(1)$. Each of the partons sees half of the gauge field, so $\pi$-flux per unit-cell, leading to a dispersion in terms of a Bloch momentum defined relative to a doubled unit-cell \cite{15_Senthil_theory,lotric_paired_2025}. The doubling of the unit-cell means that each species of parton fills one Chern $\chern=1$ band. In the doped system, each parton species has $1-2\delta$ particles and one unit flux per doubled unit cell. 

Localized anyons in the parton picture are created as composites of charge and flux by acting with a local parton annihilation operator $f_{p}(\mathbf r_0)$ \textit{and} simultaneously inserting a $\pi$-flux of the internal field $a$, with this flux being responsible for introducing the fractional quasihole exchange statistics. Indeed, we have checked that trial wavefunctions of this form accurately span the low energy Hilbert space of certain FCI models (see Sect.~\ref{sub:local_QHs} of the Supplement). Here, however, we consider systems with a finite density of delocalized anyons, and we approximate a complicated superposition of different configurations with local $\pi$-fluxes with a smeared, uniform background flux. Thus, anyons with localized fractions of flux are replaced by empty orbitals in the parton wavefunctions and a uniform background mean-field. This un-binding of the charge and the flux of the quasiholes is the crucial assumption leading to all the anyon states proposed in this work. But how do we determine the value of the uniform background flux? 
The crucial insight of \cite{senthil_doping_2024} was that if the partons are different, each partonic band reacts differently to additional flux being inserted, which means that it will generally be favourable to have a mean-field flux \textit{Ansatz} which assigns average magnetic fluxes $\pi+b$ and $\pi-b$ per unit cell with $b\neq0$, unlike the $\pi$-flux in the undoped case (the sum of the two fluxes must be fixed for the resulting bosonic state to be in the lowest Landau level, as explained in Appendix~\ref{sup:parton_flux_sum}). A particularly nice arrangement comes when $b=\pm2\pi\delta$ (take $+$ without loss of generality), which gives parton 2 a density $1-2\delta$ of particles per doubled unit cell in a density $2\pi-4\pi\delta=2\pi(1-2\delta)$ of flux, meaning it continues to fill the lowest band. In fact, $|b|=2\pi\delta$ is the largest value we can have before being forced to move partons into a higher band, which is assumed costly. Parton 1 however now has $1-2\delta$ particles in flux density $2\pi+4\pi\delta$, leading to a density $4\delta$ of quasiholes, which experience an effective magnetic field (particle-hole conjugating and noting that the flux on a lattice is only defined modulo $2\pi$) $-4\pi\delta=2\pi(-2\delta)$, meaning they likely form a $\tilde\nu=-2$ (gapped) IQHE state. 

Parton 1 couples to the internal $\U(1)$ field $a$ and consists of (1) a $\chern=1$ band and (2) the $\tilde\nu=-2$ IQHE of quasiholes, while parton 2 couples to $A-a$ and consists of only a $\chern=1$ band. Both partons are gapped (parton 1 has a gap $\sim\delta$, while parton 2 has a larger gap), and we may integrate them out to get
\begin{equation} \label{eq:bose_half_basic_sc}
    \mathcal{L}=\frac{1-2}{4\pi}ada+\frac{1}{4\pi}(A-a)d(A-a)=\frac1{4\pi}AdA-\frac1{2\pi}adA.
\end{equation}
The quadratic terms for $a$ cancel, and Eq.~\ref{eq:bose_half_basic_sc} can be shown (for example via the 3D particle-vortex duality) to describe a charged superfluid with the order parameter being the monopole operator $\mathcal M_a$ inserting a flux $2\pi$ of the field $a$~\cite{05_barkMcGreevyTheory,seiberg_duality_2016,senthil_duality_2019,PESKIN1978122,Halperin_1981_u1_duality}. We present a brief explanation of this crucial fact, which underlines all superconducting and superfluid states in this work, in Appendix~\ref{sup:cssc}. One way to see this is to observe that the equations of motion for $a$ in Eq.~\ref{eq:bose_half_basic_sc} imply $dA=0$, which is just the Meissner effect. In this way, doping the $\nu=\frac12$ bosonic FCI leads to an anyon superfluid state. While this state originates from the physics of anyons, it is adiabatically connected to the regular superfluid of the underlying bosons, as has been argued in Ref.~\cite{senthil_doping_2025}. This may also be seen by noting that we have the two partons in $\chern=\pm1$ bands, which is the same description as that discussed in the context of superfluid-FCI transitions \cite{05_barkMcGreevyTheory,lotric_paired_2025} and is the trivial superfluid.

For fermions at $\nu=\frac13-\delta$, the idea is similar. There are now three partons, but if they are all different, the flux will preferably re-arrange such that all quasiholes are associated with the same parton, say $f_3$. In particular, each parton has $1-3\delta$ particles per unit-cell, and we can expect them to feel fluxes $2\pi(1-3\delta)$, $2\pi(1-3\delta)$, $2\pi(1+6\delta)$. Partons $f_{1,2}$ remain gapped, while $f_3$ has $9\delta$ quasiholes in effective field $-12\pi\delta$, giving $\tilde\nu=-\frac32$. This is gapless at the parton mean-field level and requires more careful analysis. Ref.~\cite{senthil_doping_2024} argues that a ``secondary composite Fermi liquid (CFL)'' can form in this half-filled hole Landau level and that this CFL may undergo a pairing instability, becoming gapped at very long wavelengths. But at any rate, the situation here is more complicated.

An important result of \cite{senthil_doping_2024} came when considering the $\nu=\frac23$ fermionic Jain state, which may be understood by partonizing $\psi_f=f_1f_2f_3$ and putting partons in bands with Chern numbers $(\chern_1,\chern_2,\chern_3)=(1,1,-2)$. It was noted that if $f_3$ has the smallest gap, which corresponds to the $Q=1/3$ anyon being the cheapest excitation per unit charge, the result is similar to what we saw for $\nu=\frac13$ but if $f_{1,2}$ have the smallest gap, physically if the $Q=2/3$ anyon is cheap, the situation is similar to the $\nu=\frac12$ bosonic case in that we get a gapped parton mean-field state which corresponds to an anyon superconductor. 

A unifying interpretation of these results is that one should look at the lowest-charge local excitation that can be created out of the anyons you are doping. If that is a boson (three $Q=2/3$ anyons lead to a Cooper pair, while at $\nu=\frac12$ two $Q=1/2$ anyons make the microscopic boson), the resulting state is a superconductor. But if we make a fermion first (three $Q=1/3$ anyons), the resulting state has a Fermi surface and might later flow towards something else \cite{senthil_doping_2024,senthil_doping_2025}. This framework relies on the assumption that the partons are \textit{different} at the mean-field level and that, as a consequence, it is energetically preferred for one parton to ``carry'' all the quasiholes. But in essentially all models where a (nearly) exact ground state wavefunction is known (Landau levels and ideal/vortexable bands \cite{ledwith_vortexability_2023,wang_exact_2021}, discussed in Sect.~\ref{sub:nonuniformB}), that wavefunction has an identical mean-field \textit{Ansatz} for at least some of the partons (or in Laughlin-like cases for all the partons). With that, it is no longer obvious that this picture holds up. The main contribution of the present work is a detailed analysis of the rich physics that emerges upon doping the physically realistic Laughlin-adjacent states, where some partons are equivalent.

We start the next section by considering a Landau level with a weak periodic potential modulation and develop a framework for thinking about the energetics of doping in a parton mean-field picture. In Sect.~\ref{sub:nonuniformB}, we consider the effects of moving to more general ideal bands akin to Landau levels in non-uniform magnetic fields and in Sect.~\ref{sub:beta_int} we describe the effects of parton interactions. This mean-field theory is tested with Variational Monte Carlo in Sect.~\ref{sect:vmc}.

\subsection{Mean-field energetics of doping} \label{sub:mfdope}
As discussed in Eq.~\ref{eq:laughlin_as_partons}, Laughlin's wavefunction may be exactly factorized into $m$ equal partons. Consider now a scenario where the FQH state has a many-body gap $\Delta$, and we introduce a periodic modulation of strength $V_m\ll\Delta$ (such that there is one flux quantum per unit cell of the potential) to the system, effectively turning the lowest Landau level into an ideal Chern band with uniform Berry curvature. This will introduce a dispersive band of width $\sim V_m$ for the quasiholes and the partons.  The final energy scale in the problem is the single-particle cyclotron frequency $\omega_{c,0}\gg\Delta$, which we assume to be infinite in our analytical calculations, effectively projecting to the lowest Landau level.

Upon hole-doping, it will be favourable to remove partons from orbitals at the top of these bands. But the system may redistribute the effective flux seen by the partons to alter these bands.  This is certainly the expectation if the partons are different, but if the mean-field \textit{Ansätze} of (at least some of the) parton flavours are identical, other possibilities exist. If flux does not redistribute, each parton sees a renormalised potential $\tilde V_m$ with flux $2\pi/m$ per unit-cell. Taking an $m$-fold increased unit-cell, this is still an ideal band for small $\tilde V_m\sim V_m$. With a density $\delta\ll1$ of holes, the mean-field flux may change by $\mathcal{O}(\delta)$ --- to understand these states, we first build a picture of what happens when the parton band is subject to a weak additional magnetic field.

\subsubsection{Ideal band in a magnetic field} \label{subsub:bandinfield}
Consider a fermion in a weak periodic potential $\hat V$ where the flux per unit cell is $2\pi+bA_\text{uc}$ with $|b|\ll1$   where $A_{\text{uc}}$ is the area of the unit cell.  So we are considering a unit cell with one quantum plus a small deviation. In what follows, length units are chosen such that the area of the unit cell is $A_\text{uc}=1$.

When $b=0$, the dispersion may be found by projecting $\hat V$ into the LLL -- if $\hat V = \sum_\mathbf q V_\mathbf q \rho_\mathbf q$ (with $\rho_\mathbf{q}$ the Fourier transformed density operator), $\hat V_\text{LLL}=P\hat VP=\sum_\mathbf{q} V_\mathbf q e^{-q^2/(4B)}\bar\rho_\mathbf{q}$, where  $\mathbf{q}$ is a reciprocal lattice vector of the periodic potential.  At $b=0$, the $\bar\rho_\mathbf q$ commute and may all be simultaneously diagonalized, $\bar\rho_\mathbf q=e^{i\mathbf{q}\cdot\mathbf{R}}$ giving us a band structure $V(\mathbf R)$ in terms of the guiding-centre coordinate $\mathbf R$, which may be related to a Bloch momentum via $\mathbf R=(\hat z\wedge \mathbf k)\ell_B^2$, giving $V(\mathbf k)$.

In our problem, we are interested in nearly full parton bands, with a density $\sim\delta\ll1$ of holes near the top of the band. To understand the effects of a magnetic field, we focus on expansions around a maximum of the dispersion --- $V(\mathbf k_0+\mathbf q)=V_\text{max.}-\frac12 m^{*-1}_{\mu\nu}q^\mu q^\nu$ (for $\nu=\frac1m$, projective translations demand that each parton has $m$ degenerate maxima of this type). For $b\neq0$, like for a free particle, the quadratic dispersion gives way to mini Landau-Levels, the top of which is at $V_\text{max.}-\frac12|b|/\sqrt{\det m^*_{\mu\nu}}$. In terms of $V_\mathbf{q}$, we can write
    \begin{align} \label{eq:eff_mass_tensor_expansion}
    V(\mathbf k_0+\mathbf q)&=\sum_\mathbf p V_\mathbf p e^{-\frac{p^2}{4B}}e^{i \mathbf p\wedge (\mathbf k_0+\mathbf q)/B}\nonumber\\ &\approx V_\text{max.}-\frac12m^{*-1}_{\mu\nu} q^\mu q^\nu \\
    m^{*-1}_{\mu\nu}&=\frac{1}{B^2}\sum_\mathbf p V_\mathbf p e^{-\frac{p^2}{4B}}e^{i\mathbf p\wedge \mathbf k_0/B}(\hat z\wedge\mathbf p)_\mu(\hat z\wedge\mathbf p)_\nu \nonumber
    \end{align}

\begin{figure}
    \centering
    \includegraphics[width=0.9\linewidth]{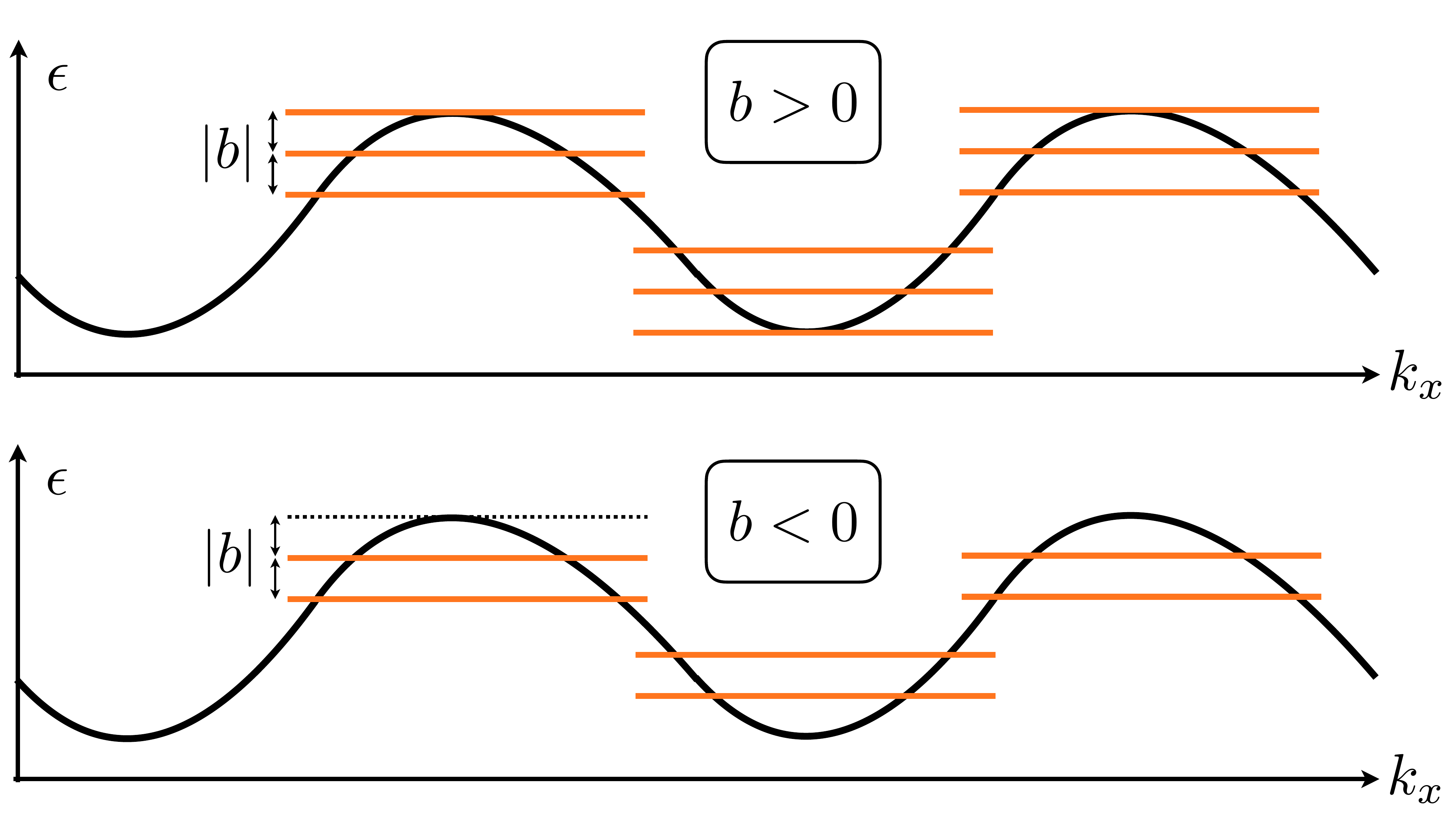}
    \caption{The formation of mini Landau Levels (mLLs) when a small magnetic field is added to an ideal Chern band. Depending on the direction of the field, the first mLL forms either exactly at the top of the band or an energy $\tilde\omega_c=|b|/m^*$ away. This is in contrast to time-reversal invariant bands, where the first mLL's forms at energy $\tilde\omega_c/2$ away from the extremum regardless of the sign of $b$. In $\chern=1$ bands, the density of states per unit cell is $(1+b/2\pi)$, unlike the $b$-independent value in the presence of time-reversal symmetry.  At $\nu=\frac1m$ with $b=0$, each parton's dispersion has $m$ degenerate maxima. When $b\neq0$, this degeneracy can be broken by inter-peak tunnelling processes, but the scale of this effect is $\sim \exp(-\frac{2\pi}{m^2}\frac1b)$, making it negligible at small $b$.
    }
    \label{fig:sketch_mlls}
\end{figure}

But the magnitude of the terms in the Landau Level-projected potential $\hat V_\text{LLL}$ also depends on the magnetic field via the projector $P$ --- in a larger field, the same physical potential leads to a stronger dispersion. We have [to lowest order, $V_\text{max.}=V(\mathbf k_0)$]
\begin{equation} \label{eq:derivative_wrt_b}
\begin{split}
    \frac{\partial}{\partial B}V(\mathbf k_0)&=\frac{1}{B^2}\sum_\mathbf p V_\mathbf p e^{-\frac{p^2}{4B}}e^{i \mathbf p\wedge\mathbf k_0/B} \left[|\mathbf{p}|^2/4 -i \mathbf p\wedge\mathbf k_0\right]=\\
    &=\frac{1}{4B^2}\sum_\mathbf p V_\mathbf p e^{-\frac{p^2}{4B}}e^{i \mathbf p\wedge\mathbf k_0/B}|\mathbf{p}|^2
\end{split}
\end{equation}
The simplification in the second line follows from the fact that $\mathbf k_0$ is a local extremum of the potential. Comparing this to Eq.~\ref{eq:eff_mass_tensor_expansion} shows that adding a small magnetic field $b$ will contribute a term $\frac b4 \text{Tr}m^{*-1}_{\mu\nu}$ to the energy of the peak of the band. Assuming that the effective mass tensor is isotropic, $m^{*-1}_{\mu\nu}=m^{*-1}\delta_{\mu\nu}$ (the lattice breaks the continuous rotational symmetry, but we can expect lowest-order expansions around a peak to be isotropic if the lattice preserves some subgroup [$C_3,~C_4,~C_6$] of the full rotation symmetry), we find that the top mini-LL is at $V_0-\frac1{2m^*} (|b|-b)$, with consecutive mLL's following with a spacing of $\tilde\omega_c=|b|/m^*$. Thus, surprisingly, for $b>0$, the energy of the top orbital is independent of $b$ to first order, as is sketched in Fig.~\ref{fig:sketch_mlls}. 

A few further comments are in order --- of course, if $b\neq0$, we lose the Bloch momentum as a good quantum number. Furthermore, if the system is put on a torus of size $N_x\times N_y$ unit cells, the band has $N_\phi=N_xN_y+N_xN_yb/2\pi$ states (the last term must be an integer, which quantizes the smallest changes in $b$ we can probe). Finally, owing to the Galilean invariance of a Landau Level, the average energy of an orbital is $\bar E=\frac{1}{N_\phi}\text{tr} P\hat V=V_{\mathbf{q}=0}$. It does not depend on $b$ and may be set to $\bar E =0$. As we discuss in Sect.~\ref{sub:nonuniformB} below and in Appendix~\ref{sup:nonuniform_berry}, most of the above argument (including the picture in Fig.~\ref{fig:sketch_mlls}) continues to hold for more general ideal bands. The only exception is that due to the lack of Galilean invariance, $\bar E$ can obtain a dependence on $b$.

\subsubsection{Doped state energy} \label{subsub:dopedstateE}
With an understanding of the behaviour of a single band, we now return to the parton mean-field. From a re-arrangement of the parton-flux, each parton $p$ will see an effective additional magnetic field $b_p$ which must strictly obey $\sum_p b_p=0$, see Appendix~\ref{sup:parton_flux_sum}. The parton $p$ will have a total of $N_xN_y(1-\delta)$ particles in $N_\phi^{(p)}=N_xN_y(1+b_p/2\pi)$ orbitals, leading to a density $\delta+b_p/2\pi$ of holes in an effective field $-b_p$ (due to opposite charge of holes), meaning they fill $\tilde\nu_p=-2\pi\delta/b_p-1$ mini Landau levels. Due to the projective action of the translations on the partons, $T_xT_yT_x^{-1}T_y^{-1}=e^{2\pi i/m}$, there are $m$ degenerate band maxima and mLL's form in each of these ``valleys'' for each parton. Within each ``valley'', the $n$-th mLL is at energy $V_\text{max.}-\frac{1}{2m^*}[(1+2n)|b|-b]$ and there are $m$ mLL's at that energy across the valleys.

\begin{figure}
    \centering
    \includegraphics[width=0.98\linewidth]{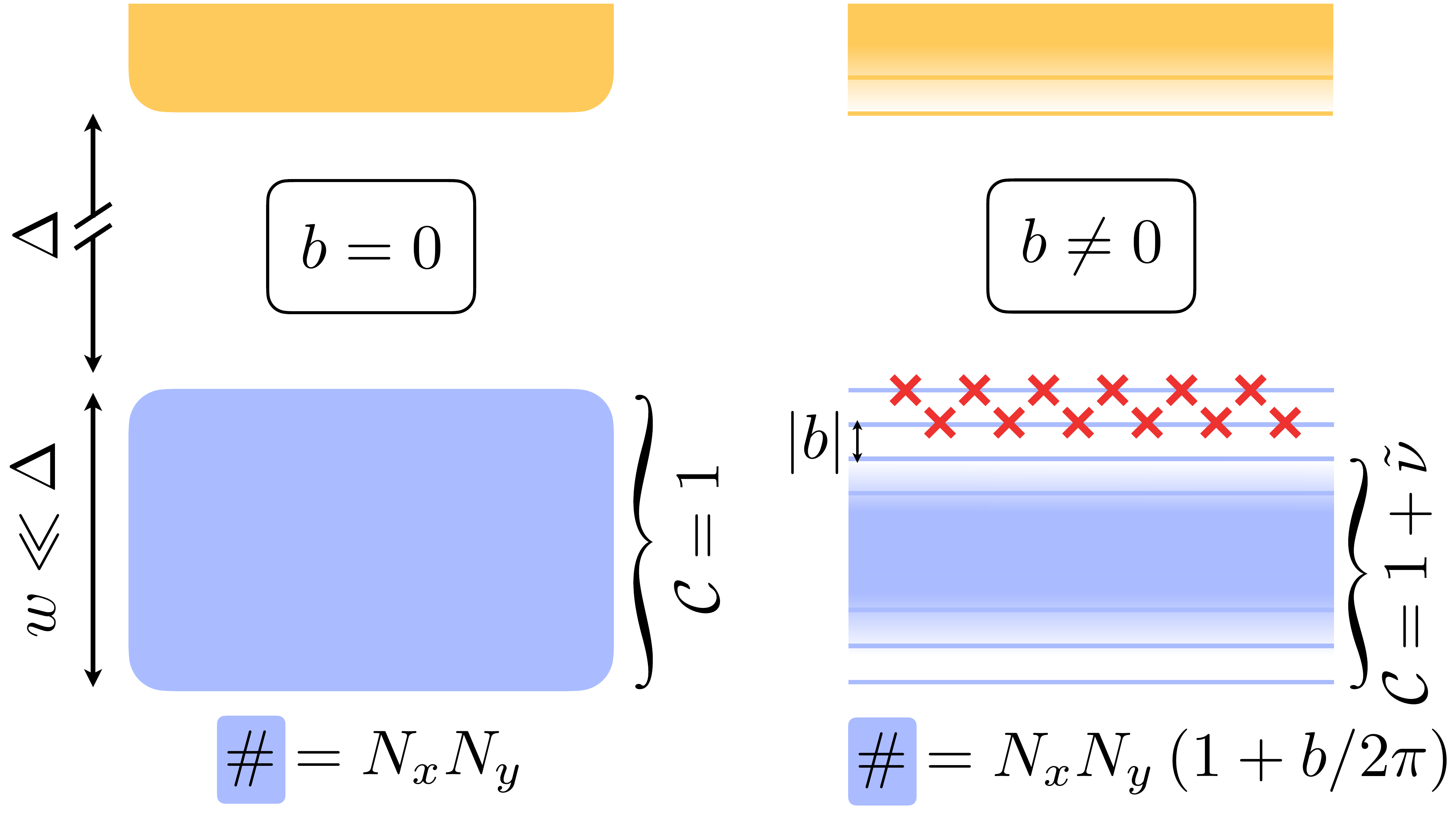}
    \caption{The mean-field parton energy levels before and after introducing a flux $b$. Initially, we have a flat $\chern=1$ band (blue) of width $w\ll\Delta$ with $\Delta$ the parton band gap  (which is at the scale of the physical many-body gap). Upon introducing a small field $b$, the flat-band splits into mLL's, spaced by an energy $\sim |b|$. The band now has $N_xN_y(1+\frac b{2\pi})$ states. Given a fixed doping, for sufficiently negative $b$, some partons must jump the gap $\sim\Delta$, which is large. The lower band will generally not be filled, and if the holes (red crosses) are at mLL filling $\tilde\nu$, the remaining partons fill a $\chern=1+\tilde\nu$ band with a gap $\sim |b|$ and an enlarged unit cell of area $\sim 2\pi/|b|$.}
    \label{fig:partonband_energyscales}
\end{figure}

Decomposing $|\tilde\nu_p|/m=n_p+r_p$ into the integer and fractional parts (with $n_p=\lfloor|\tilde\nu_p|/m\rfloor$), the total energy density of such a parton band is (after particle-hole conjugating around $\bar E=0$) 
\begin{align} \label{eq:mf_energetics_full_uniform}
    E_p = &-\left(\delta+\frac{b_p}{2\pi}\right)V_\text{max.} + \sum_{k=0}^{n_p-1} \frac{m|b_p|}{2\pi m^*}\left[(1+2k)|b_p|-b_p\right] \nonumber\\
    &+ r_p\frac{m|b_p|}{2\pi m^*}[(1+2n_p)|b_p|-b_p].
\end{align}
The mean-field state will minimize $E=\sum_pE_p$ given the constraints that $\sum_pb_p=0$ and that $b_p\geq-2\pi\delta$ for each $p$. If this latter constraint is violated, there exists a parton for which the number of states in the lower band after flux re-distribution, $N_\phi^{(p)}$, is smaller than the number of particles. This requires excitations out of the lowest parton band, which are assumed to be costly, as illustrated in Fig.~\ref{fig:partonband_energyscales}.

\begin{figure}
    \centering
    \includegraphics[width=0.95\linewidth]{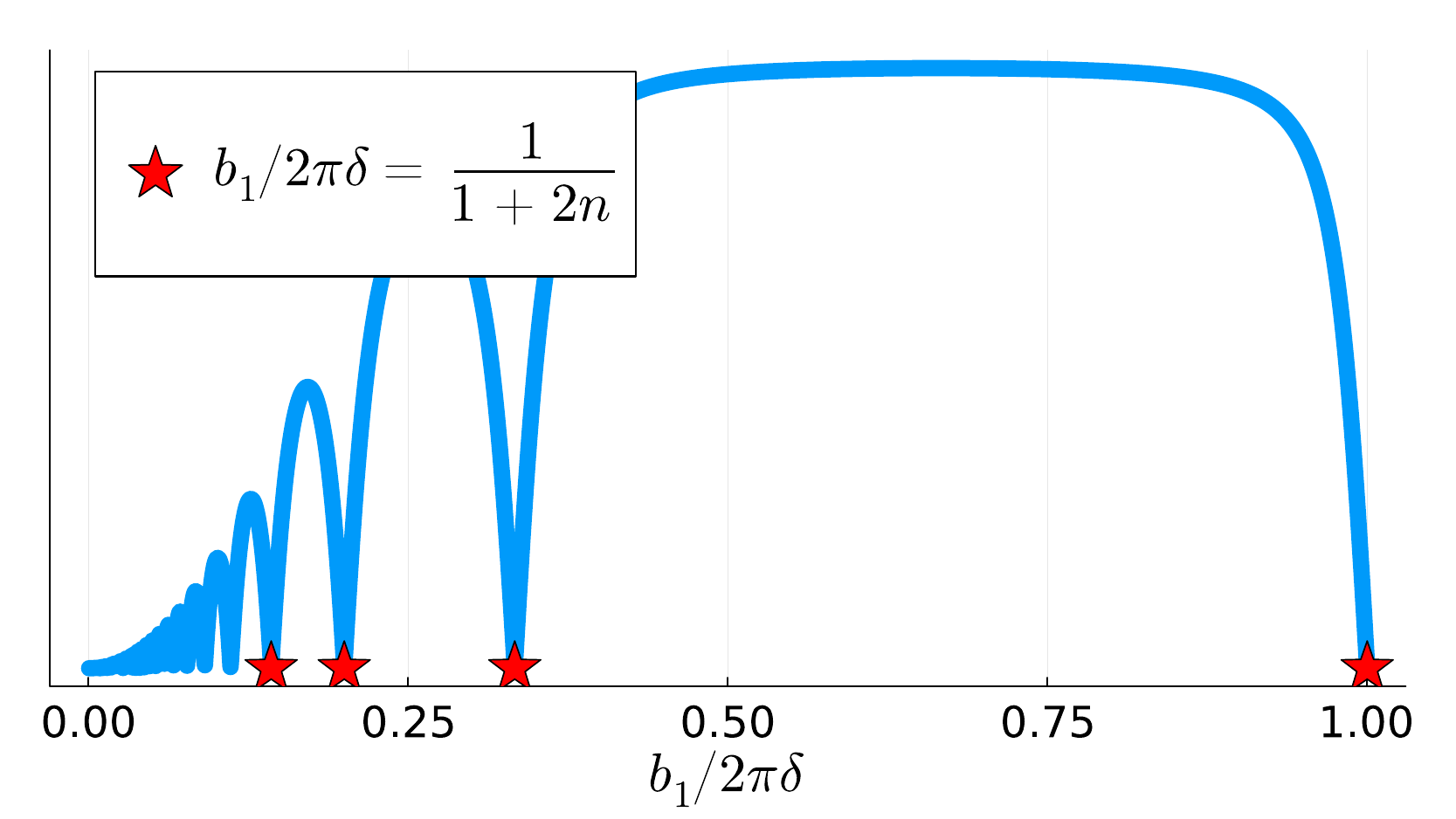}
    \caption{The parton mean-field energy of a doped $\nu=\frac12-\delta$ FCI as a function of the flux $b_1$ of the internal gauge field, showing degenerate minima whenever the partons are gapped at the mean-field level, at $b_1=2\pi\delta /(1+2n)$ for all $n\in\mathbb Z_{\geq0}$.}
    \label{fig:bosonic_mf_line}
\end{figure}

To build intuition, consider first the case of $\nu=\frac12$ bosons, partonized as $b=f_1f_2$. We may parametrize $b_1=2\pi\delta x,~b_2=-2\pi\delta x$ and take $x\in[0,1]$ to describe the degree of flux re-distribution ---  if the parton flavours have differing mean-field \textit{Ansätze} before doping, $x=1$ is expected after doping, but now we consider models where the parton mean-field \textit{Ansätze} are equal before doping. We plot the resulting doped energy from Eq.~\ref{eq:mf_energetics_full_uniform} as a function of $x=b_1/2\pi\delta$ in Fig.~\ref{fig:bosonic_mf_line}.
The result indicates degenerate minima for all values of $x=\frac{1}{2n+1}$ with $n\in\mathbb{Z}_{\geq 0}$ --- at those values, the parton filling fractions are $\tilde\nu_1=-2-2n,~\tilde\nu_2=2n$ --- they each fill $-(n+1)$ and $n$ full mLL's in each of the $m=2$ valleys. The states that fill mLL's in both valleys for both partons all have the same energy and this is the lowest energy of any state. The partons in these states are also gapped with a gap $\sim|b_p|\sim \delta/(2n+1)$. We may integrate them out while remembering the two nearly full parton bands in the $\nu=\frac12$ state to give the effective Lagrangian (each parton fills a $\chern_p=1+\tilde\nu_p$ band with a bandgap $\sim|b_p|$, see Fig.~\ref{fig:partonband_energyscales}) 
\begin{equation}
\begin{split}
    \mathcal L_{x=\frac{1}{2n+1}}&=\frac{1+\tilde\nu_1}{4\pi}ada+\frac{1+\tilde\nu_2}{4\pi}(A-a)d(A-a)\\&=-\frac{1+2n}{2\pi}adA+\frac{1+2n}{4\pi}AdA
    \end{split}
\end{equation}
which describes a charge-$(1+2n)$ superfluid resulting from a partial flux-redistribution upon doping a system of charge-1 bosons (See Appendix~\ref{sup:cssc}). The case $n=0$ reduces to the prior work (See Sect.~\ref{sub:senthil_cstheory} and Ref.~\cite{senthil_doping_2024}), but this construction demonstrates the possibility of other states, which are all degenerate at this level of analysis. We should not expect this degeneracy to hold when treating general Hamiltonians beyond the mean-field level --- in Sect.~\ref{sub:nonuniformB}, we identify the interplay between Berry curvature and effective dispersion in an ideal band as a tuning parameter selecting which states are favoured, and in Sect.~\ref{sub:beta_int} we discuss how interactions can also pick out specific states.

\begin{figure}
    \centering
    \includegraphics[width=0.9\linewidth]{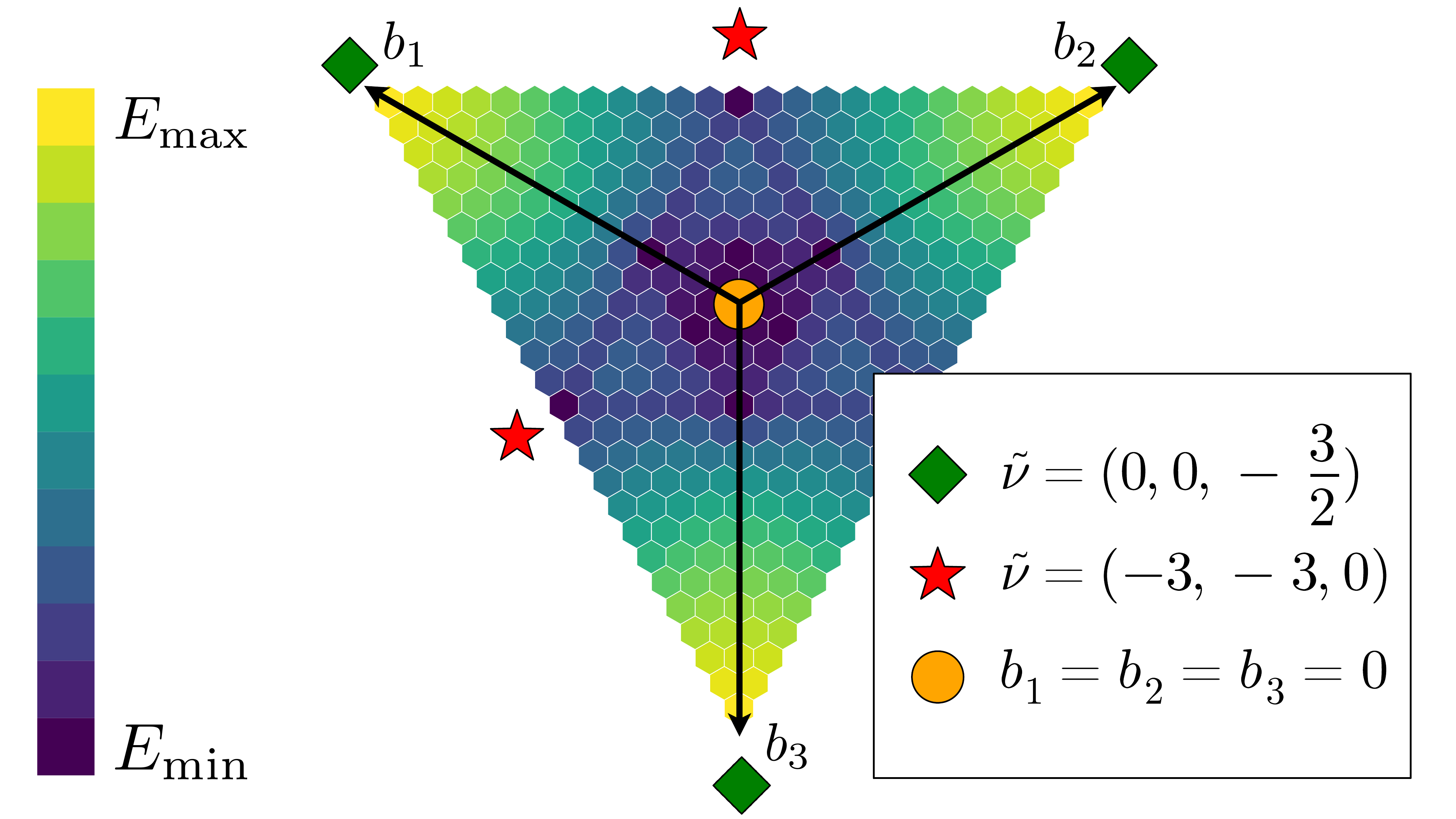}
    \caption{Parton mean-field energy of a doped $\nu=\frac13-\delta$ FCI. The centre (orange dot) indicates a state with $b_1=b_2=b_3=0$ where all partons are equivalent and see no magnetic field, while the corner states (green diamonds) are cases where $b_1/2=-b_2=-b_3=2\pi\delta$ and only one of the partons holds quasiholes. A two-dimensional continuum of possible flux arrangements spanned by $b_1,b_2,b_3$ with the constraint $b_1+b_2+b_3=0$ lies between these extremes. The filling fractions are related to the fields by $\tilde\nu_p=-1-\frac{2\pi\delta}{b_p}$ and evolve non-linearly across the diagram, diverging at the origin.}
    \label{fig:fermi_mf_hexagon}
\end{figure}

The case of the $\nu=1/3$ Laughlin state is more interesting --- we have discussed in Sect.~\ref{sub:senthil_cstheory} that in the case of different parton mean-field \textit{Ansätze}, the predicted mean-field state is not gapped and is likely not a superconductor. But is a superconductor possible when the partons are equal? We now have three fluxes $b_{1,2,3}$ --- with the constraint $\sum_pb_p=0$, this gives a two-dimensional space of possible fluxes, with no configuration favoured a priori by one parton being inherently different. Eq.~\ref{eq:mf_energetics_full_uniform} may be used to predict the energy of different flux arrangements, and the results are shown in Fig.~\ref{fig:fermi_mf_hexagon}.
The outcome is somewhat similar to the bosonic case --- again Eq.~\ref{eq:mf_energetics_full_uniform} predicts that states where all $\tilde\nu_{1,2,3}$ are integer multiples of $m=3$ are the best energetically --- these are also the only states with a gap at the mean-field parton level and they are all degenerate. But note that, unlike in the $m=2$ case, here the fully polarized corners (which are similar to the states that would be realized if the partons were different) do not fully fill mLL's and are not deemed energetically favourable. The red stars in Fig.~\ref{fig:fermi_mf_hexagon} indicate one of the gapped states with (up to a permutation) $\tilde\nu=(-3,-3,0)$. Among the many degenerate gapped states, this one is special as it has the largest single-parton gap. 
 Furthermore, the partons $f_{1,2}$ share an identical mean-field \textit{Ansatz} in this state, which leads to a large, $\U(2)=[\SU(2)\times \U(1)]/\mathbb Z_2$ gauge invariance of the parton theory ---  this is a larger gauge group than the $\U(1)\times \U(1)$ expected for general flux arrangements, but still a subgroup of the high-energy $\SU(3)$ gauge invariance. To write an effective Chern-Simons theory, we denote the $\U(1)$ part of the gauge field by $a$ and let the partons be charged under $(A-a,A-a,2a-A)$, with the doublet $(f_1,f_2)$ additionally transforming in the fundamental representation of a gauge field $\alpha\in \text{su}(2)$. Remembering that the partons effectively fill gapped bands with a gap $\sim\delta$ and Chern numbers $\chern=1+\tilde\nu=(-2,-2,1)$ (Fig.~\ref{fig:partonband_energyscales}), we get
\begin{equation} \label{eq:one_third_u2_sc_state}
\begin{split}
    \mathcal L =& -\frac{2}{4\pi}\text{Tr}\left[\alpha d\alpha + \frac23 \alpha^3+(A-a)d(A-a) \mathbb 1_2\right]\\
    &+ \frac{1}{4\pi}(2a-A)d(2a-A)=\\
    =&-\frac2{4\pi}\text{Tr}\left[\alpha d\alpha+\frac23\alpha^3\right]-\frac{3}{4\pi}AdA+\frac{2}{2\pi}adA.
\end{split}
\end{equation}
The coupling to the external electromagnetic field is only via $\frac{2}{2\pi}adA$, meaning that the final term leads to a charge-$2e$ superconductor, with the monopole of the gauge field $a$ playing the role of an order parameter,  as reviewed in Appendix~\ref{sup:cssc}. In Eq.~\ref{eq:one_third_u2_sc_state}, we observe that the $\SU(2)$ component $\alpha$ does not couple to $A$. It will only leave a signature in the quantized thermal Hall conductance, $\kappa_{xy}=\frac{\pi^2 k_B^2 T}{3h} c_-$, where Eq.~\ref{eq:one_third_u2_sc_state} predicts
a central charge $c_-=-3+3/2=-3/2$. The first term in this expression comes from the $n=-3$ IQHE response $-\frac{3}{4\pi}AdA$ and the second arises from the charge-neutral non-Abelian $\SU(2)_{2}/\mathbb Z_2$ theory present \footnote{References \cite{senthil_doping_2025,shi2026charge4esuperconductorparafermionicvortices} demonstrate that this superconductor is related to the $p+ip$ chiral superconductor.} We thus expect the $\tilde\nu=(-3,-3,0)$ state to be a chiral superconductor of half-integral central charge $c_-=-3/2$. This is a significant theoretical prediction --- assuming the weak pairing regime (See~\cite{46_Read_Green}), this corresponds to an odd real-space pairing symmetry, which in turn implies the triplet electron spin pairing channel. This is consistent with the fact that our model takes spin-polarised electrons (i.e. spinless fermions) as an input and should be contrasted with the prediction of integral central charge in Ref.~\cite{senthil_doping_2024}, which while not entirely inconsistent with triplet pairing (as a direct correspondence between pairing symmetry and central charge only exists in the weak-pairing regime of \cite{46_Read_Green}, but not in all cases) seems less natural.  Recently, Ref.~\cite{shi_non-abelian_2025} has proposed a theory analogous to Eq.~\ref{eq:one_third_u2_sc_state} in the context of bandwidth-tuned transitions out of an FCI and Ref.~\cite{shi2026charge4esuperconductorparafermionicvortices} has further analysed properties of similar superconducting constructions arising out of forms of $\U(2)$ Chern-Simons theory. Our work differs from these in two significant aspects --- (1) we obtain this state by doping the FCI, which is the more experimentally relevant scenario and (2) we do not only propose Eq.~\ref{eq:one_third_u2_sc_state} as a possible effective theory of some state, but our work for the first time also proposes concrete physical Hamiltonians (in Sect.~\ref{sub:nonuniformB}--~\ref{sub:parton_mf_summary}) under which such a state is expected to be energetically favourable. In Sect.~\ref{sect:vmc} we confirm that it is a viable ground state for the doped FCI problem. The $\tilde\nu=(-3,-3,0)$ state is not the only superconductor in Fig.~\ref{fig:fermi_mf_hexagon} and in Sect.~\ref{sup:cs_general_sc} of the Supplement, we demonstrate that any gapped doped state of that form where $\tilde\nu_{1,2,3}$ are all integers leads to superconductivity.

Particle-hole conjugating the fractionally filled band allows for an identical construction at $\nu=\frac23+\delta$, when particle-doping the $\nu=\frac23$ FCI. In that situation, which has been studied numerically \cite{wang_chiral_2025}, particle-hole conjugation gives our state a central charge $\tilde c_-=1-c_-=5/2$, while the pairing symmetry found using DMRG on a somewhat similar model in Ref.~\cite{wang_chiral_2025} would suggest $\tilde c_-=-1/2$, a point to which we return in Sect.~\ref{sub:compare_23_dmrg}.  This prediction of how a half-integer central charge superconductor emerges out of doping the $\nu=\frac13$ or $\nu=\frac23$ FCIs is a crucial result of our work.

 This anyon superconductor goes against the picture laid out in Sect.~\ref{sub:senthil_cstheory} (and Refs.~\cite{senthil_doping_2024,senthil_doping_2025}) that the way to get a superconductor would be to dope in the $Q=\frac23$ anyon. The field-theoretical result in Eq.~\ref{eq:one_third_u2_sc_state} certainly demonstrates that this state is a superconductor, but we would ideally want a complementary, more physical picture of why \textit{six} $Q=\frac13$ anyons in this state bunch into Cooper pairs instead of electrons (as \textit{triplets} of anyons). In Sect.~\ref{sup:stack_and_condense} of the Supplementary, we attempt a discussion of how this state could be understood within the algebraic theory of anyons, but much of that picture remains unclear.

\subsubsection{Feedback effects from parton non-uniformity}
If each parton species fills a Landau Level, we obtain a state with full Galilean translational invariance, but this is not the case for doped states --- when some mLL's are filled with holes, the density $n_p=\langle f_p^\dagger f_p\rangle$ will maintain the translational invariance with the unit-cell of $V(\mathbf x)$, but partons will be preferably depleted from the part of the unit-cell with $V(\mathbf x)$ large. Any non-uniformity of $n_p(\mathbf x)$ changes the effective gauge field and background potential felt by the other partons, which could modify our picture significantly. But crucially, the average field felt by the partons is $m$-fold smaller than the microscopic field, making the parton effective magnetic length larger than the lattice length by a factor $\sim \sqrt m$. For large $m$, this means that the lattice spacing is too small for the density to vary appreciably without paying a kinetic energy cost, leading to an approximately uniform parton mean-field state. For a square lattice, we show in Appendix~\ref{sup:flux_redist} that such feedback effects are suppressed in powers of $\epsilon=e^{-\pi m/2}$ for the gapped states where all $\tilde\nu_p/m \in \mathbb Z$ when compared to the energies of Eq.~\ref{eq:mf_energetics_full_uniform}. Already for $m=2,3$, the effect of this on picking which of the gapped states is favoured is generally negligible compared to the band-geometry and interaction-based mechanisms we turn to now.

\subsection{Beyond Landau Levels} \label{sub:nonuniformB}
A powerful generalization of the above calculation (Section~\ref{subsub:dopedstateE}) is to allow for the magnetic field to be non-uniform within a unit-cell while keeping the total flux per unit cell the same. The resulting bands cover a wide class of band structures, including ``ideal'' and the more general ``vortexable'' bands \cite{wang_exact_2021,ledwith_family_2022,ledwith_vortexability_2023,estienne_ideal_2023}. These bands may have both a non-trivial dispersion and a non-uniform Berry curvature (unlike the uniform curvature in a Landau Level), but they do satisfy some version of a ``trace condition'' on their Quantum Geometric Tensor (QGT) which helps with analytical progress.  We work with ideal bands, noting that any vortexable band may be converted into an ideal one by a choice of unit-cell embedding~\cite{ledwith_vortexability_2023, Simon_2020}. In fact the parton numerical work in Sect.~\ref{sect:vmc} is indifferent to the embedding and sees no distinction between the two notions.
It is generally believed that the bands of twisted MoTe$_2$ are well approximated by such ideal bands. The full QGT technology is not required for this calculation, however, and it is sufficient to simply think about a construction analogous to Sect.~\ref{subsub:dopedstateE}, with some (possibly large) modulation of the magnetic field added \cite{estienne_ideal_2023,wang_exact_2021}. So we consider our system in a ``Generalized Landau level'' (GLL) with a non-uniform magnetic field.

We parametrize the position-dependent field in an ideal band as  $B(\mathbf{r})=B_0+\nabla^2K(\mathbf{r})$ with $B_0A_\text{uc}=2\pi$ and with $K$ having lattice translational invariance with respect to our unit cell. It may be shown that the band is spanned by the wavefunctions $\ket{\psi_k}$
\begin{equation} \label{eq:ac_band_wfs_form}
    \braket{\mathbf{r}|\psi_k}=e^{-K(\mathbf{r})}\braket{\mathbf{r}|\psi^\text{LLL}_k},
\end{equation}
so simply the LLL, multiplied by an orbital-independent factor.  Crucially, for an ideal flat band and for short range repulsive interactions, the ground state may be shown to be related to the Laughlin state, $\psi(\mathbf{r}_1,\ldots,\mathbf r_N)=e^{-\sum_j K(\mathbf{r}_j)}\psi^\text{Laughlin}(\mathbf{r}_1,\ldots,\mathbf{r}_N)$ \cite{wang_exact_2021,ledwith_vortexability_2023}. 

The $\nu=\frac1m$ Laughlin state may be exactly factorized into $m$ equal partons as in Eq.~\ref{eq:laughlin_as_partons}, all filling the LLL at a reduced field. Similarly, this state may be factorized exactly into $m$ equal partonic factors, now each filling an ideal band with an $m$-fold larger unit-cell and magnetic field $\tilde B=B_0/m + \nabla^2K/m$. Because the partons are all equal, we can expect the physics of Sect.~\ref{sub:mfdope} to play an important role. We note that at every real-space coordinate $\mathbf x$, the effective magnetic fields felt by the partons $B_p(\mathbf x)$ must sum to the total external magnetic field $B(\mathbf x)=\sum_p B_p(\mathbf x)$ for the resulting parton wavefunction of the form Eq.~\ref{eq:general_parton_to_wf} to be in the lowest Landau level, a point which we prove in Appendix~\ref{sup:parton_flux_sum}.

The crucial difference compared to Sect.~\ref{sub:mfdope} is in the particle-hole conjugation of parton bands --- whereas before, we had $\bar E=\text{Tr} P \hat V/N_\phi=0$ for any field (the band is always centered at zero energy), this ceases to be the case here, as a non-zero $K(\mathbf r)$ leads to modulations in the real-space density $n_p(\mathbf r)$ and this modulation depends on the additional flux inserted. Say that a parton sees a small flux $b$ different from that in the undoped case. Then, the particle-hole conjugation around a filled parton band contributes an energy per particle $\bar E=\gamma b + \eta b^2+\ldots$. Noting that the partons see fluxes $b_p$ with $\sum_pb_p=0$, we get that the total energetic contribution will be $E'=\eta\sum_p b_p^2$. In Appendix~\ref{sup:nonuniform_berry}, we show that the rest of the argument in Sect.~\ref{sub:mfdope} for uniform fields carries over to this case. The energy $E'$ can break the degeneracy of states seen in Figs.~\ref{fig:bosonic_mf_line},~\ref{fig:fermi_mf_hexagon} and favour either more flux polarization ($\eta<0$) or less flux polarization ($\eta>0$).

Finding $\eta$ analytically for a general ideal band is difficult, but we can make progress assuming that the magnetic flux is only weakly non-uniform --- we set $K(\mathbf{r})=\kappa M(\mathbf{r})$ and work to first order in $\kappa$.  Eq.~\ref{eq:ac_band_wfs_form} gives a wavefunction basis for the band, and a useful object to define is the projector onto this generalized Landau Level (GLL), $P^\text{GLL}=\ket{\psi_k}G^{-1}_{kq}\bra{\psi_q}$ with $G_{kq}=\braket{\psi_k|\psi_q}$ the Gram matrix. Given this GLL projector and a single-particle potential $V$, we may infer the number of orbitals and average energy as $N_\phi=\text{Tr}(P^\text{GLL})$ and $N_\phi \bar E=\text{Tr}(P^\text{GLL}V)$.

To first order, we have $G_{kq}=\delta_{kq}-2\kappa \bra{\psi^\text{LLL}_k}{M}\ket{\psi_q^\text{LLL}}$, thus $G^{-1}_{kq}=\delta_{kq}+2\kappa \bra{\psi^\text{LLL}_k}{M}\ket{\psi_q^\text{LLL}}$, from which we get the projector
\begin{equation}
\begin{split}
    P^\text{GLL}=& ~e^{-\kappa M}\ket{\psi_k^\text{LLL}}G^{-1}_{kq}\bra{\psi_q^\text{LLL}}e^{-\kappa M}=\\
    =& ~e^{-\kappa M}\ket{\psi^\text{LLL}_k}\bra{\psi^\text{LLL}_k}e^{-\kappa M}\\
    & +2\kappa \ket{\psi^\text{LLL}_k}\bra{\psi^\text{LLL}_k} M\ket{\psi^\text{LLL}_q}\bra{\psi^\text{LLL}_q}=\\
    =& ~P-\kappa MP -\kappa PM + 2\kappa PMP
\end{split}
\end{equation}
where $P=\ket{\psi^\text{LLL}_k}\bra{\psi^\text{LLL}_k}$ is the LLL projector. It follows (from $MV=VM$ and trace cyclicity) that 
\begin{equation}
    N_\phi\bar{E}=\text{Tr}(P^\text{GLL}{V})=2\kappa\left[-\text{Tr}(P{V}{M})+\text{Tr}({V}P{M}P)\right],
\end{equation}
which expresses the average energy of an orbital in the non-uniform field entirely in terms of the LLL wavefunctions and the modulations $M(\mathbf r),V(\mathbf r)$. For a magnetic field $B_0$, recall that  $|\braket{\mathbf{r}|P|\mathbf{r'}}|^2=\frac{B_0}{2\pi}e^{-B_0|\mathbf r - \mathbf r'|^2/2}$, which leads us to
\begin{align}
    N_\phi\bar E=&-2\kappa\int \text{d}^2\mathbf r \frac{B_0}{2\pi}V(\mathbf r)M(\mathbf r)\\
    &+2\kappa\int\text{d}^2\mathbf r\text{d}^2\mathbf r' \left[\frac{B_0}{2\pi}\right]^2V(\mathbf r)M(\mathbf r')e^{-B_0|\mathbf r-\mathbf r'|^2/2} \nonumber
\end{align}
Now expanding $B_0=2\pi+b$, we get $\bar E=\bar E_0+\gamma b+\eta b^2+\dots$ with
\begin{equation} \label{eq:modification_nonuniform_field}
\begin{split}
\gamma =&~ -\frac\kappa\pi \int\text{d}^2\mathbf r V(\mathbf r)M(\mathbf r)\\
&-\frac{\kappa}{2\pi} \frac{1}{N_\phi}\int\text{d}^2\mathbf r\text{d}^2\mathbf r' V(\mathbf{r})H(\mathbf{r}')e^{-B_0|\mathbf r-\mathbf r'|^2/2}, \\
\eta =&~ \frac{\kappa}{2\pi B_0}\frac{1}{N_\phi}\int\text{d}^2\mathbf r\text{d}^2\mathbf r' V(\mathbf r) H(\mathbf r')G(\mathbf{r}-\mathbf{r'}),
\end{split}
\end{equation}
where $H(\mathbf{r})=\nabla^2M/B_0=\frac{1}{\kappa}\frac{B(\mathbf{r})-B_0}{B_0}$ represents the local deviation from uniformity of the magnetic field ($H(\mathbf{r})\sim\mathcal{O}(1)$ as $\kappa$ has been factored out), and $G(\mathbf{r})=(\frac12+\frac{B_0|\mathbf r|^2}4)e^{-B_0|\mathbf r|^2/2}$ is a normalized kernel.

The parameter $\eta$ is thus determined by an interplay of the fluctuations in the effective parton magnetic field (related to fluctuations in Berry curvature~\cite{wang_exact_2021}) and the fluctuations of $V$, which determines the effective parton dispersion. Note that even if the underlying physical band is entirely flat, the combination of particle interactions and the non-uniformity of the magnetic field still generates an effective parton/quasihole dispersion, as has been recently discussed in detail by Ref.~\cite{yan_anyon_2025}. Our analysis stems from assuming nearly uniform Berry curvature, but it demonstrates that the interplay between the fluctuations in the effective magnetic field $H(\mathbf r)$ and the background potential $V(\mathbf r)$ can function as a tuning knob. This allows us to choose whether $\eta\sim\int VH>0$, discouraging large $|b_p|$ or $\eta\sim\int VH<0$, encouraging large $|b_p|$, although the precise effect is more difficult to compute for highly non-uniform magnetic fields.

\subsection{The effect of interactions} \label{sub:beta_int}

Another effect that might discriminate between the different states proposed in Sect.~\ref{sub:mfdope} (Figs.~\ref{fig:bosonic_mf_line},~\ref{fig:fermi_mf_hexagon}) are inter-particle interactions. The effect here is subtle and is discussed in some detail in Sect.~\ref{sup:interactions_effect} in the Supplement. The main argument comes from the fermionic nature of the partons --- because each parton $f_p$ is a spinless fermion, Pauli's exclusion principle lowers the amplitude for two parton holes of the same species to approach each other. From this, we can reason that at the mean-field level, parton $p$ has a density $\rho_p=\delta+b_p/2\pi$ of quasiholes and the total quasihole density is $\sum_p\rho_p=m\delta$. We expect that a pair of quasiholes associated with the same parton is less likely to approach each other than a pair associated with different partons.
Assuming a local interaction term $\sim U$ (with $U>0$ meaning repulsive interactions), we should thus expect $E_\text{diff}-E_\text{same}\sim U$. There are $\frac12(\sum_p\rho_p)^2=\frac12m^2\delta^2$ total quasihole pairs, out of which $\frac12\sum_p \rho_p^2$ are of the ``same'' type and the remaining are of the ``different'' type. The energy difference is thus $\frac12E_\text{same}\sum_p\rho_p^2+\frac12E_\text{diff}(\sum_{q\neq p}\rho_p\rho_q)=E_0-\frac{U}{4\pi^2}\sum_p b_p^2$, where we used $\sum_p b_p=0$. The physically more realistic repulsive interactions $U>0$ tend to favour larger $\sum_p b_p^2$. Combining this with the band geometry effects means that an additional term $(\eta_\text{geo.}+\eta_\text{int.})\sum_pb_p^2$ is to be added to Eq.~\ref{eq:mf_energetics_full_uniform}, which we derived for non-interacting electrons in a Landau level in Sect.~\ref{sub:mfdope}. The crucial fact that we use in the following is that we may choose a particular form of the repulsive interaction (giving some fixed $\eta_\text{int.}<0$) and still use the geometric contribution $\eta_\text{geo.}$ to control which states are favoured, with the interactions only re-normalizing the point at which the system changes behaviour.

In fact, recently, Ref.~\cite{yan_anyon_2025} has carefully studied the behaviour of anyons in an FCI in a perfectly flat ideal band with repulsive Coulomb interactions, finding an effective dispersion $\epsilon(\mathbf k)$ for the quasiholes. If we imagine this dispersion as being generated by a smooth applied $V^\text{eff}(\mathbf r)$, their results indicate that $V^\text{eff}(\mathbf r)$ is small in the region of positive $(B(\mathbf r)-B_0)$, implying that $\int V^\text{eff}(\mathbf r)(B(\mathbf r)-B_0)<0$, which leads to the band geometry considered there having $\eta<0$. But this is purely an interaction effect, and it is consistent with the picture we give that repulsive interactions contribute a negative term to $\eta$.

\subsection{Summary of parton mean field results} \label{sub:parton_mf_summary}

To summarize, we predict that the equality of partons leads to a competition between many candidate doped states. Depending on the interactions and the band geometry, doping the $\nu=\frac12$ bosonic FCI can either lead ($\eta<0$ in Eq.~\ref{eq:modification_nonuniform_field}) to the anyon superconductor or  ($\eta>0$) to a state where the partons are gapless at the mean-field level and where $\SU(2)$ gauge invariance is maintained --- we discuss this state in Sect.~\ref{sect:undistr} after a numerical confirmation of our theory in Sect.~\ref{sect:vmc}. For the $\nu=\frac13$, predicted states include the ``secondary CFL'' of Ref.~\cite{senthil_doping_2024} (for $\eta\ll0$), a superconductor with half-integer central charge for $\eta\lesssim 0$ and an $\SU(3)$ mean-field-gapless parton state for $\eta>0$.

\section{Variational Monte Carlo results} \label{sect:vmc}
The parton construction has allowed us to develop some picture of the energetics of these doped states, but we should remember that even in the case of a gapped parton mean-field, the approximation is still highly uncontrolled. One way to move beyond the pure parton mean field is to recall that each parton mean-field state $\ket\Omega$ implies a microscopic wavefunction (See Eq.~\ref{eq:partonform}), which may be compared to the true ground state, as we have previously done in Ref.~\cite{lotric_paired_2025}.   The idea here is to constrain ourselves to parton mean-fields of hole-doped states with a particular arrangement of the effective parton fluxes $b_p$ and find the lowest energy state in that sector --- optimizing over all possible noninteracting parton wavefunctions in Eq.~\ref{eq:partonform} corresponding to that flux. In practice, this is done by demanding that the effective parton mean-field Hamiltonian $H^{(p)}_\text{MF}$ obeys lattice translational invariance $[H^{(p)}_\text{MF},T_{x/y}^{(p)}]=0$ with projective translation operators $T_x^{(p)}T_y^{(p)}=e^{i\phi_p}T_y^{(p)}T_x^{(p)}$. This guarantees that the parton sees flux $\phi_p=2\pi/m+b_p$ per unit cell. Comparing the best variational energies between different $b_p$ clarifies which arrangement is preferred, at which point we can use the effective Chern-Simons theory to find what phase this corresponds to. To improve the performance of our \textit{Ansatz}, we allow for an additional \textit{real} Jastrow factor $\prod_{i<j}e^{-J(\mathbf r_i,\mathbf r_j)}$ in front of our wavefunction with $J$ translationally invariant and $J\neq0$ only when $|\mathbf r_i-\mathbf r_j|$ is sufficiently small. We expect this Jastrow factor to not affect the phase of our system, only changing the microscopic energetics to improve the variational energy. More details about the VMC procedure are given in Appendix~\ref{sup:vmc}.

We work in an exact lattice mapping of the LLL, following the work by Kapit and Mueller \cite{kapit_exact_2010}. We generalize the model to not only exactly represent the Landau level of uniform flux, but to instead work for any $B(\mathbf r)=B_0+\nabla^2K$. The function $K$ is periodic with respect to translations by $d_x,d_y$ original unit cells, which sets the larger, $d_x\times d_y$ unit cell of the ideal band. The field is set such that exactly one unit of flux passes through each $d_x\times d_y$ cell. The construction is detailed in Appendix~\ref{sup:vmc}. A dispersion may be added to the band by introducing a different chemical potential $V(\mathbf r)$ for the different sites in the $d_x\times d_y$ cell while respecting our theory's assumption of the energy scale hierarchy $V\ll\Delta\ll\omega_{c,0}$.

In what follows, we choose $d_x=d_y=3$ and define the two periodic functions to take on values $V(\mathbf r)=V_0[\cos(2\pi \mathbf r \cdot\hat x)+\cos(2\pi \mathbf r \cdot\hat y)]$ and $K(\mathbf r)=K_0[\cos(2\pi \mathbf r \cdot\hat x)+\cos(2\pi \mathbf r \cdot\hat y)]$ for sites positioned on $\mathbf r=(l_x/d_x,l_y/d_y)$ with $l_{x,y}=1\ldots d_{x,y}$ indexing the location within the unit-cell. 

Working in energy units where the band gap of this model is unity, $\omega_c=1$, we first note that purely hard-core interactions are enough to stabilize the bosonic FCI with a many-body gap $\Delta=0.05$, observed in exact diagonalization on a small system. For the fermionic case, we choose a nearest-neighbour interaction of strength $U=6$ which yields a many-body gap $\Delta=0.15$. We expect the gaps to remain similar for the system sizes we consider. To control the anyon dispersion, we introduce the periodic modulation of strength $V_0=0.0075$. After projecting this modulation to the lowest band, it yields a dispersion and an effective width $w\sim V_0 e^{-\pi/2}\sim0.001$ of the band. This arrangement of scales gives $w\ll\Delta\ll\omega_{c,0}$, as we assumed when developing the theoretical picture. The value of $K_0$ may be used to tune between the phases. From $B(\mathbf r)-B_0=\nabla^2K=-(2\pi/d_{x,y})^2 K$, we get that $\eta_\text{geo.}\propto -V_0K_0$ and so the sign and magnitude of $K_0$ can be used to control the behaviour of the system.

\begin{figure}
    \centering
    \includegraphics[width=0.99\linewidth]{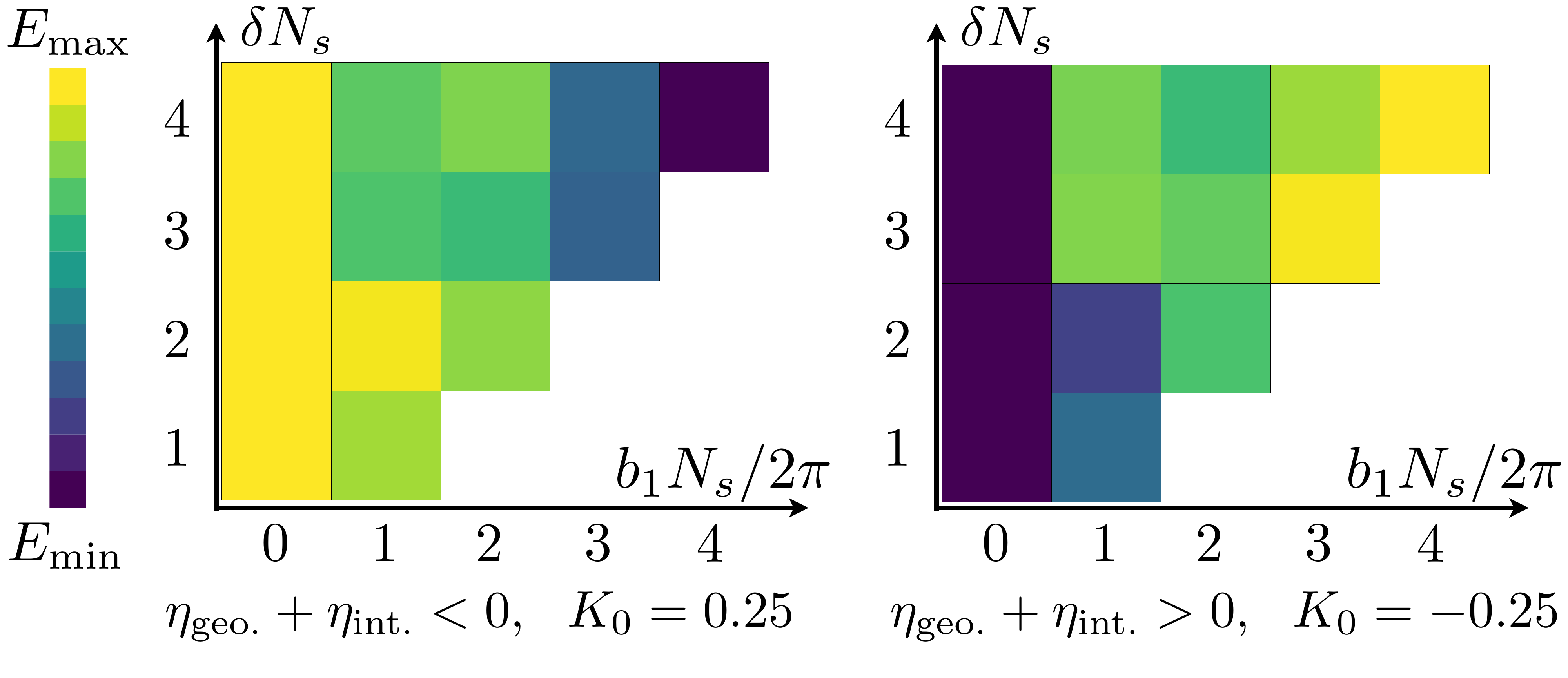}
    \caption{Bosons at $\nu=\frac12-\delta$ on a system of $N_s=48$ unit cells ($6\times8$ grid) for $\delta\in [0.02,0.08]$. The nature of the ground state depends on the sign of $\eta=\eta_\text{geo.}+\eta_\text{int.}$ (as interactions are hard-core only, $\eta_\text{int.}=0$). When $\eta<0$, the lowest energy state has $b_1=2\pi\delta$ and forms an anyon superconducting state with $\U(1)$ gauge invariance, as predicted by \cite{senthil_doping_2024}. When $\eta>0$ however, $b_1=0$ is favoured for all $\delta$, leading to a gapless parton mean-field with an $\SU(2)$ gauge invariance which is discussed further in Sect.~\ref{sect:undistr}.     }
    \label{fig:boson_results}
\end{figure}

Focusing first on bosons at $\nu=\frac12-\delta$, the results in Fig.~\ref{fig:boson_results} largely confirm the theoretical predictions. As a result of the two partons generally being different, we have only one $\U(1)$ gauge field whose saddle point flux we can vary. 
When $\eta<0$ ($K_0=0.25$ { in Fig.~\ref{fig:boson_results}}), we find that the mean-field parton flux of the state with the lowest energy changes by one unit for each additional boson we remove. This corresponds to $b_1=2\pi\delta$ which, recalling our discussion in Sect.~\ref{sub:senthil_cstheory}, leads to one parton remaining gapped without quasiholes and the other seeing a $\tilde\nu=-2$ hole IQHE. The resulting state is a regular unit-charge bosonic superfluid. When we tune to $\eta>0$ ($K_0=-0.25$ 
{ in Fig.~\ref{fig:boson_results}}), however, the situation changes. Now regardless of the number of removed particles, we find that $b_1=0$ has the lowest energy. In this case, both partons remain gapless and the $\SU(2)$ gauge invariance is preserved in the low-energy theory. Determining the resulting phase is not trivial due to the gapless partons, but we make some comments on the likely nature of it in Sect.~\ref{sect:undistr}. 

\begin{figure*}
    \centering
    \includegraphics[width=0.99\linewidth]{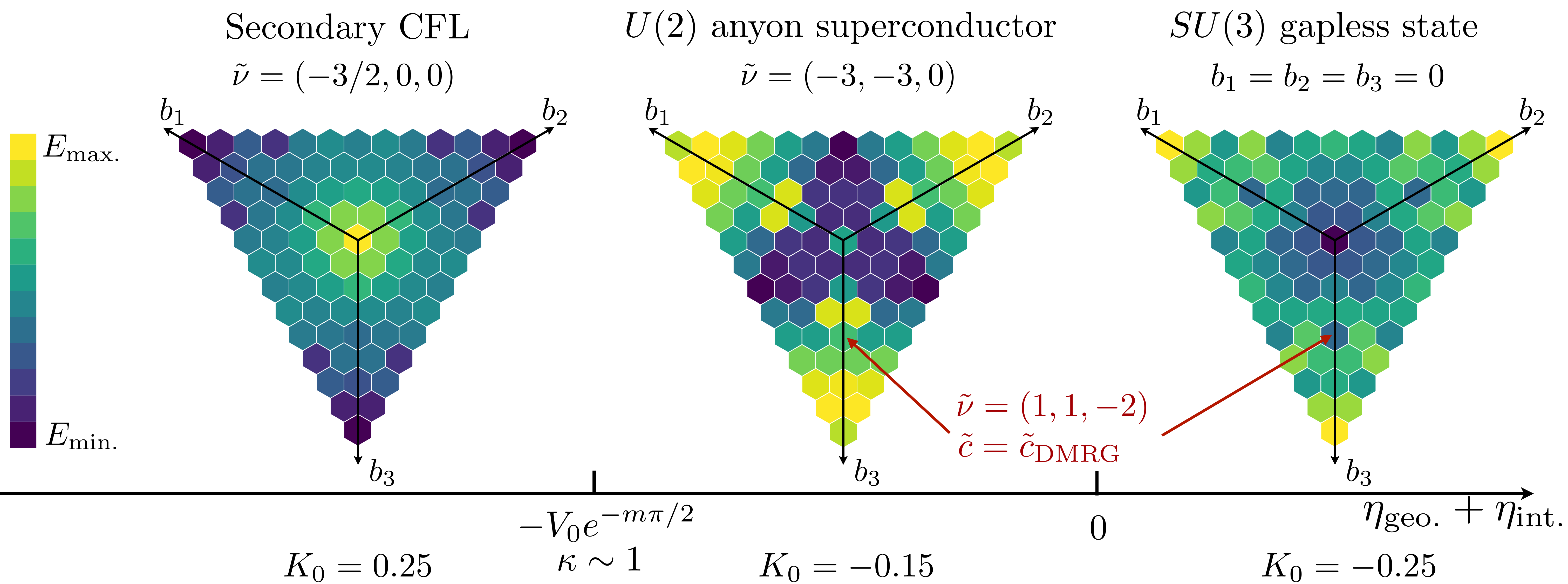}
    \caption{Energies from doping the fermionic FCI at $\nu=\frac13-\delta$ with $\delta=1/9$ as a function of the mean-field flux seen by the partons. Depending on $\eta=\eta_\text{geo.}+\eta_\text{int.}$, the ground state may either be the secondary composite Fermi liquid as predicted in Ref.~\cite{senthil_doping_2024}, a type of anyon superconductor (discussed in Sect.~\ref{sub:mfdope}) or a gapless state that preserves the full $\SU(3)$ gauge invariance, discussed in Sect.~\ref{sect:undistr}. The competition is determined by a combination of particle interactions and the non-uniformity of the magnetic field $B(\mathbf r)$. For a fixed particle interaction, the intermediate anyon superconducting state is observed over an $\mathcal O(1)$ range of the effective field non-uniformity $\kappa \sim \max|B(\mathbf r)-B_0|/B_0$. The results were obtained on a $6\times6$ torus. We point out the existence of an anyon superconducting state with $\tilde\nu=(1,1,-2)$ and $\U(2)$ gauge invariance where the predicted chiral central charge is consistent with the pairing symmetry of the DMRG results in Ref.~\cite{wang_chiral_2025}, although that state is not found favourable in our calculations and likely breaks translational symmetry at the parton mean-field level.    
    }
    \label{fig:fermions_results}
\end{figure*}

The case of fermions  $\nu=\frac13-\delta$ displays richer structure. With three partons $\psi_f\sim f_1f_2f_3$, we have at least a $\U(1)\times\U(1)$ gauge invariance at the mean-field level or maybe $\U(2)$ or $\SU(3)$. The numerical results for $\delta=\frac19$ are shown in Fig.~\ref{fig:fermions_results} and the outcome may be understood within the mean-field theory developed in Sect.~\ref{sub:mfdope}. Starting at $\eta<0$ by letting $K_0=0.25$, we find that the energetically favoured state is the fully polarized Secondary CFL predicted in Ref.~\cite{senthil_doping_2024}. This is to be expected, as a large negative $\eta$ will pick out the state with the largest $\sum_p  b_p^2$.  Decreasing $K_0$, we find a different behaviour at $K_0=-0.15$, where the lowest-energy state has $\tilde\nu=(-3,-3,0)$.  This is the $\U(2)$ anyon superconductor of half-integer central charge we discussed in some detail in Sect.~\ref{sub:mfdope}, so the VMC results indeed confirmed that this anyon superconducting state is a viable ground state. The appearance of this state here may be explained by the fact that out of all the states that are degenerate at the mean-field level in the absence of quantum geometry and interactions (in Fig.~\ref{fig:fermi_mf_hexagon}), this one has the largest $\sum_p b_p^2$, which means that it should be the favoured state when $\eta<0$ with $|\eta|$ small. While it may be surprising to find the $\tilde\nu=(-3,-3,0)$ state at $\eta_{\text{geo.}}\propto -K_0>0$, this re-normalization of the transition point is entirely consistent with the expectation that repulsive interactions introduce an additional $\eta_\text{int}<0$ as discussed in Sect.~\ref{sub:beta_int}. Finally, at $K_0=-0.25$, we find that the lowest energy state is the point with $b_1=b_2=b_3=0$ and $\SU(3)$ gauge invariance.  As with the analogous case in the bosonic FCI, the partons are gapless, and the fate of these states is a challenging question to which we turn to in Sect.~\ref{sect:undistr}. Finally, we present results of an analogous calculation at $\delta=\frac{1}{18}$ in Sect.~\ref{sub:smallerdope} of the Supplement, confirming the important features of the results in Fig.~\ref{fig:fermions_results}.

\subsection{Relation to the $\nu=\frac23$ state} \label{sub:compare_23_dmrg}
Our work has focused on doping $\nu=\frac13$ as this state offers a simple trial wavefunction description. By particle-hole conjugation, any state we constructed here for $\nu=\frac13-\delta$ may be used to propose a state at $\nu=\frac23+\delta$, which is a situation that has been studied numerically in Ref.~\cite{wang_chiral_2025}. After particle-hole conjugation, we have argued in Sect.~\ref{subsub:dopedstateE} that our $\U(2)$ anyon superconductor has central charge $\tilde c_-=5/2$, in conflict with the $\tilde c_\text{DMRG}=-1/2$ expected based on the pairing symmetry within DMRG. One important caveat is that the symmetry of the order parameter can, but does not have to, agree with the chiral central charge (see, for example, Ref.~\cite{senthil_doping_2024,shi_non-abelian_2025,wang_chiral_2025}).  We point out that among the states we considered, there exists at least one other charge-$2e$ anyon superconductor with a $\U(2)$ gauge invariance, the one obtained at $\tilde\nu=(1,1,-2)$, also indicated in Fig.~\ref{fig:fermions_results}.  This state shares most of its properties with the $\tilde\nu=(-3,-3,0)$ case, but the parton Chern numbers are $\chern=1+\tilde\nu=(2,2,-1)$, the opposite of the situation in $\tilde\nu=(-3,-3,0)$ where $\chern = 1 + \tilde\nu =(-2,-2,1)$. This means that the thermal Hall responses of the two states are time-reversal conjugates, leading to $c_{(1,1,-2)}=3/2$ for this state and after particle-hole conjugating it to form an \textit{Ansatz} at $\nu=\frac23+\delta$, we find a central charge $\tilde c_{(1,1,-2)}=1-c_{(1,1,-2)}=-1/2$ in agreement with the DMRG results. In our calculations, however, this state is energetically unfavourable both in the mean-field analysis in Sect.~\ref{sub:mfdope} and in the VMC results. One reason for this is that since $\tilde\nu$ are not integral multiples of $3$, our analysis suggests that the state cannot be simultaneously translationally invariant and gapped at the mean-field level. In the case that translational symmetry is broken to introduce a gap, the state is still a superconductor, see Sect.~\ref{sup:cs_general_sc} of the Supplement. But it may be argued that at least the particle density might only break the translational invariance to $\mathcal O(\delta^2)$ despite the naive expectation of $\mathcal O (\delta)$ effects (Sect.~\ref{sup:cs_general_sc} of the Supplement). It remains a question whether this non-uniformity remains visible after including gauge field fluctuations in the calculation.  Furthermore, we stress that the microscopic details of the model considered in Ref.~\cite{wang_chiral_2025} are strictly different from our model, which could also be the reason for the disagreement.

\section{Effective theory for the gapless $\SU(m)$ states} \label{sect:undistr}
Our newly discovered states with $\eta>0$ warrant more discussion. We start by considering the bosonic case, where to zeroth order, each parton species has (due to the projective symmetry) two Fermi surfaces. All four resulting Fermi surfaces are coupled by an $\SU(2)$ gauge field, since parton flavour symmetry is unbroken. In $\SU(2)$ gauge theory, the interaction in the singlet $f_1f_2$ channel is attractive (while in the triplet channel it is repulsive), suggestive of a pairing instability with $\langle f_1f_2\rangle$, see \cite{senthil_doping_2025} or Sect.~\ref{sup:su_m_csm} of the Supplement.

But there exist two ways of gapping the state via pairing since each parton has two Fermi surfaces (FS's). Say that the LLL-projected microscopic potential  $V_\text{LLL}(\mathbf R)$ has a maximum at some $\mathbf{R}_0$ --- this implies that the bosonic dispersion maximum will be at $\mathbf k_0 = B(\mathbf R_0\wedge\hat z)$ and that the parton dispersion maxima (the FS's) are at (remembering the partons see field $B/2$) $\mathbf k_{\mathbf G}=(B/2)(\mathbf{R}_0+\mathbf a)\wedge\hat z=\mathbf k_0/2+\mathbf G/2$ with $\mathbf G$ a reciprocal vector of the microscopic (bosonic) unit cell. Pairing like maxima between $f_{1,2}$ gives a bosonic order parameter at momentum $2\mathbf k_\mathbf{G}=\mathbf k_0$, matching the location of the peak in the bosonic dispersion, where we expect to first see effects upon removing a small fraction of bosons, although it is possible that for some microscopic models, the alternative pairing pattern which gives a charged superfluid at some $\mathbf k_0+\mathbf G/2$ (three different such points exist in the microscopic BZ). Our current numerical technique is not capable of making accurate statements about parton behaviour beyond the mean-field level, but explorations of how a fluctuating Chern-Simons field may be added to the procedure could prove useful in this context.
We thus expect the undisturbed bosonic state to also be a superfluid, but this one resulting from a pairing instability of a parton Fermi surfaces of size $k_F\sim\sqrt\delta$ instead of as a consequence of fluctuations in the flux of a $\U(1)$ gauge field via the usual ``anyon superconductivity'' mechanism.

A similar question may be asked about the fermionic state at $b_1=b_2=b_3=0$ (the ground state for large enough $\eta>0$). We have three partons with $(f_1,f_2,f_3)$ transforming in the fundamental rep. of an internal $\SU(3)$ field --- as all partons are equal, we need to consider $\SU(3)$ fluctuations. The gauge field mediated interactions favour states where groups of three partons together form $\SU(3)$ singlets $f_1f_2f_3\sim \psi_f$ (Sect.~\ref {sup:su_m_csm} of the Supplement). Following a similar line of reasoning to the bosonic case, we denote by $\mathbf k_0$ the peak of the non-interacting electron dispersion, which implies parton dispersions have peaks at $\mathbf k_0/3+\mathbf G/3$. Combining partons from minima with the same $\mathbf G$ into singlets gives a Fermi surface of microscopic electron holes at $\mathbf k_0$. So for $\eta>0$, we would expect to observe the Laughlin state, stacked with a small Fermi surface of regular, non-fractionalized quasiholes which do not see the internal $\SU(3)$ field. As with the bosonic case, our analysis is too crude to determine whether this might prefer to form at a different momentum $\mathbf k_0+\mathbf G/3$. It is also in principle possible for pairs of quasiholes to first combine into the anti-fundamental rep. as $\phi=(f_2f_3,f_3f_1,f_1f_2)$ before triplets of $\phi$ condense into a superconductor. More advanced numerical techniques, or approaches treating quasiholes as the primary degrees of freedom (See \cite{yan_anyon_2025}) might be needed to understand these states better.

\section{Discussion} \label{sect:discussion}
The main takeaway of our work should be that doping FCIs can have a richer structure with a competition between more states than what has been predicted in previous work (see Refs.~\cite{senthil_doping_2024,senthil_doping_2025,sahay_superconductivity_2024,kim_variational_2025,soejima_anyon_2024,song_doping_2021,wen_topological_2025,schleith_anyon_2025}), with this structure resulting from the partons all being in similar mean-field states in the FCI. Alternatively, we may phrase this as treating the Laughlin states as $\SU(m)_1$ CS theory instead of the more typical dual $\U(1)_m$ --- while equivalent in the undoped case, we have seen that the predictions the two theories make when doping are different.  We stress that states with parton symmetry are far from only being special fine-tuned points in parameter space. One prominent example are states derived from a flux attachment/composite fermion approach, where $\psi(z_1,\ldots ,z_N)=\prod_{i<j}(z_i-z_j)^{P~}\psi^\text{CF}(z_1,\ldots z_N)$. Any such state can be thought of as partonizing $\psi=f_1f_2\ldots f_{P} f_\text{CF}$, where each of the partons $f_p$ is responsible for one factor of $\prod_{i<j}(z_i-z_j)$. This form of flux attachment thus introduces $P$ flavours of partons, all of which are symmetrical, which means that these states have a large $\U(P)$ [or depending on the form of $f_\text{CF}$ even $\SU(P+1)$] gauge invariance and should display physics similar to what we discussed in this work upon doping. In light of this, the states with some degree of parton symmetry seem to be the more generic case, at least for fermionic states, where $P\in 2\mathbb Z$ guarantees at least a $\U(2)$ gauge invariance. Despite this, such parton symmetry is often ignored in similar constructions. While this discussion here is specialized to a Landau Level, the argument holds for any ideal or vortexable band where flux-attachment of this form is a well-defined operation \cite{ledwith_vortexability_2023}, meaning that the physics of symmetric partons is likely present in many microscopic models at a variety of filling fractions.

In the doped FCI problem we considered, we found that a combination of interactions, band dispersion and band geometry is responsible for determining the phase of the anyonic system. In particular, we have identified (i) a scenario where one recovers the states that would be predicted if one started from a mean-field with different partons (a $\U(1)^{(m-1)}$ picture), (ii) a scenario where the doped phase preserves the full $\SU(m)$ gauge invariance and where the parton mean-field is gapless and (iii)  an intermediate scenario where doping the $\nu=\frac13$ FCI with $Q=\frac13$ anyons can lead to a superconducting state. The $\tilde\nu=(-3,-3,0)$ superconducting state we found to be energetically favourable in this regime carries half-integer central charge $c_-=-3/2$ and is adiabatically connected to one of the proposed superconducting states in Ref.~\cite{shi_non-abelian_2025}, where it was proposed as a possible path to superconductivity upon increasing the width of the initially flat lowest band to drive the system out of an FCI. The generality of our \textit{Ansatz} also allows one to construct a similar anyon superconducting state at $\tilde\nu=(1,1,-2)$, which could be in agreement with the pairing symmetry found in Ref.~\cite{wang_chiral_2025}, although our numerical results never find that state to be favourable.

Upon hole-doping the $\nu=\frac13$ Laughlin state, one expects to generate the fundamental $Q=\frac13$ anyons. The established picture is that since three of these anyons together can form the microscopic electron, one would expect to see a state with a Fermi surface (which may then undergo other instabilities). If instead it is favourable to create $Q=\frac23$ anyons, the picture changes --- it is impossible to combine these into a single microscopic electron, as combining three gives the Cooper pair. By this mechanism, one expects to see a superconducting state at finite doping density. 
The states we discovered provide a counterexample to this argument --- while all the anyons we dope have $Q=\frac13$, details of the band dispersion and geometry can be such that the anyon motion leads to a superconductor even if that is not the only option.  Furthermore, this superconductor has a half-integer central charge, as is expected for the effectively spinless electrons considered in our model.

To connect our work to experiments, note that \cite{xu_signatures_2025} studied an FCI at $\nu=-\frac23=-1+\frac13$ and that the superconducting dome was found at $\nu=-\frac23-\delta = -1+(\frac13-\delta)$ (with $\delta>0$), meaning that our results are of direct relevance in that regime. In fact, the superconductivity was observed around $\delta\in[0.05,0.1]$, close to the doping fraction $\delta\sim 0.1$ considered in our work --- in Sect.~\ref{sub:smallerdope} of the Supplement, we show that $\delta\sim 0.05$ yields similar results. A more detailed analysis of the competition between doped anyons states for realistic models of MoTe$_2$ would be a beneficial further direction.

\section*{Acknowledgments}
T. L. acknowledges funding from Leverhulme Trust International Professorship grant (No. LIP-202-014) and S. H. S. acknowledges support from EPSRC Grant No. EP/X030881/1.

{
\appendix

\section{The constraint on the sum of parton fluxes} \label{sup:parton_flux_sum}

In our analysis, we rely on the rule that the sum of the effective magnetic fields observed by the partons $B_p(\mathbf r)$ must sum exactly to the externally applied magnetic field at every point in the system, $B(\mathbf r)=\sum_p B_p(\mathbf r)$, for the final variational wavefunction to be in the lowest Landau level (assuming each parton chooses a wavefunction entirely in its own effective LL). This section aims to provide a proof of this statement, which we do both by using parton mean-field Hamiltonians and also by considering how parton wavefunctions multiply to form a wavefunction for the microscopic particle. This result is to be understood as both a constraint on the allowable low-energy parton mean-field configurations and also as a guarantee that for all parton configurations satisfying this constraint, the microscopic kinetic term is minimized.

We always work in the Lowest Landau level --- in a magnetic field $B(\mathbf r)=\nabla\wedge\mathbf A$, this is realized as the zero mode of the Hamiltonian $H=\frac{1}{2m^*}\Pi_-\Pi_+$ where $\Pi_\pm=p_x-A_x\pm i(p_y-A_y)$, and the lowest Landau level limit may be understood as taking $\psi$'s effective mass $m^*\rightarrow0$ while keeping the other terms in the Hamiltonian (interaction, $V(\mathbf r)$, \ldots) finite. Consider now $m$ species of partons $f_1,\ldots,f_m$, each of which is in the Lowest Landau level of a Hamiltonian $H^{(p)}=\frac{1}{2m_p}\Pi_-^{(p)}\Pi_+^{(p)}$ where $\Pi^{(p)}_\pm=p_x\pm ip_y - (A_x^{(p)}\pm iA_y^{(p)})$ and $B_p=\nabla\wedge \mathbf A_p$ is the field felt by the parton. 
Being a zero mode of $\Pi_+^{(p)}$, each parton satisfies $(p_x\pm ip_y)f_p=(A_x^{(p)}\pm iA_y^{(p)})f_p$. For the microscopic particle $\psi$, we then have, using the chain rule,
\begin{equation}
\begin{split}
    \Pi_+\psi=&[(p_x+ip_y)f_1]f_2\ldots f_m + f_1[(p_x+ip_y)f_2]\ldots f_m\\
    &+\ldots +f_1f_2\ldots[(p_x+ip_y)f_m]\\
    &-(A_x+iA_y)f_1f_2\ldots f_m=\\
    =&-\left[A_x-\sum_p A_x^{(p)} +i\left(A_y-\sum_p A_y^{(p)}\right)\right]\psi\stackrel{!}{=}0,
\end{split}
\end{equation}
where we used that the partons are zero modes of $\Pi_+^{(p)}$ to get the second line. A $\psi$ constructed in this way is a zero mode of $\Pi_+$ (and consequently a zero mode of $H$) if and only if $\sum_p \mathbf A^{(p)}=\mathbf A$, so only if the gauge fields seen by the partons sum together to the external gauge field. In the lowest Landau level limit $m^*\rightarrow0$, the energy cost associated with violations of this constraint is infinite --- as a consequence, this constraint on the sum of total parton fluxes is strictly satisfied throughout our analysis.

The same calculation may be repeated at the level of trial wavefunctions --- let $B(\mathbf r)=\nabla^2K(\mathbf r)$ be the externally applied field and $B_p(\mathbf r)=\nabla^2K_p(\mathbf r)$ be the effective fields felt by the partons (note that in most of the paper, we let $B(\mathbf r)=B_0+\nabla^2\tilde K$. In this case, we may simply set $K=\tilde K+B_0|\mathbf r|^2/4$ to bridge the gap between the two, as no translational symmetry is imposed on $K$). The lowest-parton-Landau-level orbitals of parton $p$ are spanned by $\varphi_p(\mathbf r)=g_p(z)e^{-K_p(\mathbf r)}$ where $z=x+iy$ with $\mathbf r=(x,y)$ and $g_p(z)$ is a holomorphic function. From Eq.~\ref{eq:partonform}, we know that the wavefunction for $\psi$ is the product of parton wavefunctions, so to satisfy the kinetic constraints, we demand that $\varphi(\mathbf r)=\varphi_1(\mathbf r)\varphi_2(\mathbf r)\ldots\varphi_m(\mathbf r)$ be in the lowest Landau level of the external field $B(\mathbf r)$ for all possible combinations of parton orbitals. This condition directly implies that the many-body wavefunction in Eq.~\ref{eq:partonform} will also minimize the kinetic term. We explicitly have
\begin{equation} \label{eq:flux_sum_constraint_wflevel}
\begin{split}    
    \varphi(\mathbf r)&=\varphi_1(\mathbf r)\varphi_2(\mathbf r)\ldots\varphi_m(\mathbf r)=\\
    &=\prod_{p=1}^m[g_p(z)]e^{-\sum_p K_p(\mathbf r)}\stackrel{!}{=}g(z)e^{-K(\mathbf r)},
\end{split}
\end{equation}
where the expressions of the form on the right span the microscopic Landau level and $g(z)$ is a holomorphic function. The product $\prod_pg_p(z)$ is clearly a holomorphic function, but Eq.~\ref{eq:flux_sum_constraint_wflevel} enforces the non-trivial constraint that $\sum_pK_p(\mathbf{r})=K(\mathbf r)+\log f(z)$ in order to be in the LLL where $f(z)=g(z)/\prod_pg_p(z)$ is a holomorphic function. From holomorphicity, $\nabla^2\log f(z)=0$ and it follows that we need $\sum_pB_p(\mathbf r)=B(\mathbf r)$ to strictly hold for all $\mathbf r$ to not violate the LLL condition.

This calculation makes no assumptions about the nature of the magnetic field, but the result will be physically applicable only in cases where the parton decomposition $\psi=f_1\ldots f_m$ is a sensible mean-field starting point, so close to the Laughlin phase. This calculation does not strictly demand that the parton mean-field \textit{Ansätze} should all be equal in the undoped case, although that may be inferred based on the form of Laughlin's wavefunction. The results of this section apply both to cases where some (or all) of the partons are in identical mean-field states, and also to cases where all the partons are different.

\section{Superconductivity from Chern-Simons terms} \label{sup:cssc}
In the field-theoretical calculations of our paper, we routinely refer to the fact that theories of the form
\begin{equation} \label{eq:general_cs_sc_form}
    \mathcal L= \frac{Q}{2\pi}adA+\frac{k}{4\pi}AdA,
\end{equation}
where $a$ an internal $\U(1)$ gauge field to be integrated over, describes a superfluid/superconductor of charge $Q$ under the external field $A$. While this fact is well known in the literature, we find it useful to present some justifications of this fact here. We present three arguments to build intuition for why this is the case. Before doing so, we comment on the Chern-Simons term for the external field, $\frac{k}{4\pi}AdA$, which usually leads to a quantized Hall conductance. First, if $\frac{k}{2Q}\in\mathbb Z$, we may make the redefinition $a\rightarrow \tilde a=a+\frac{k}{2Q}A$, at which point Eq.~\ref{eq:general_cs_sc_form} becomes $\mathcal{L}=\frac{Q}{2\pi}\tilde a dA$. If $\frac{k}{2Q}$ is not an integer, the same transformation would violate the compactness of $\tilde a$, but this should not matter in the system's linear response to $A$, which is by itself enough to argue in favour of superconductivity. We may thus set $k=0$ in what follows without loss of generality, although we note that when computing the thermal Hall conductance and the chiral central charge (as done for the $\tilde\nu=(-3,-3,0)$ state in Sect.~\ref{subsub:dopedstateE}), one has to be more careful.

The simplest suggestion of superconducting behaviour is to note that $a$ appears only linearly in Eq.~\ref{eq:general_cs_sc_form}. Parameters appearing linearly in the action are Lagrange multipliers enforcing a constraint, in this case that $dA=0$, which translates to $\nabla\wedge \mathbf A=0$, that there must be zero flux of the external field through the system. This flux expulsion is the Meissner effect characteristic of a superconductor.

Another argument was presented in Refs.~\cite{05_barkMcGreevyTheory,senthil_duality_2019}, which noted that Eq.~\ref{eq:general_cs_sc_form} is only the lowest order in an effective expansion in the gauge field $a$, which is inevitably obtained upon integrating out the partons. We should thus also expect a Maxwell-type term $\frac{1}{2g^2} \mathfrak f\wedge\star \mathfrak f$ in the Lagrangian with $\mathfrak f=da$. The pre-factor of this term can be perturbatively expected to depend on the inverse parton gap, which is in our states in turn linked to the doping density $\delta$, leading us to expect $g^2\sim|\delta|$. We can write
\begin{equation}
    \mathcal L = \frac{Q}{2\pi}\epsilon^{\mu\nu\rho}A_\mu\partial_\nu a_\rho - \frac{1}{2g^2}(\partial_\mu a_\nu-\partial_\nu a_\mu)^2
\end{equation}
and define $\xi^\mu=\epsilon^{\mu\nu\rho}\partial_\nu a_\rho$. We can move from integrating over $a$ to $\xi$ provided that we introduce the Lagrange multiplier $\varphi$ enforcing $\partial_\mu\xi^\mu=0$, so a term $\frac{1}{2\pi}\varphi \partial_\mu\xi^\mu=-\frac{1}{2\pi}\xi^\mu \partial_\mu \varphi \subset\mathcal L$, where this choice of normalization allows for periodicity in $\varphi\leftrightarrow\varphi+2\pi$, giving $\varphi$ the interpretation of a phase \cite{senthil_duality_2019}. We get
\begin{equation} \label{eq:cssc_l_with_constraint}
    \begin{split}        
    \mathcal{L}=&\frac{Q}{2\pi}\xi^\mu A_\mu - \frac{1}{2\pi}\xi^\mu \partial_\mu\varphi - \frac{1}{2g^2}\xi_\mu \xi^\mu\\
    =&-\frac{1}{2g^2}\left[\xi_\mu-\frac{g^2}{2\pi}(QA_\mu-\partial_\mu \varphi)\right]^2 \\
    &+ \frac{g^2}{8\pi^2}(QA_\mu-\partial_\mu \varphi)^2.
    \end{split}
\end{equation}
Upon integrating over $\xi$, which gets rid of the term in the middle row, we are left with $\mathcal L\sim g^2 (QA_\mu-\partial_\mu\varphi)^2$, which is exactly the electromagnetic response of a charged superfluid where $\varphi$ is the phase of the order parameter (e.g., in a superconductor $\Delta=|\Delta |e^{i\varphi}$), consistent with the fact that $\varphi$ is only defined modulo $2\pi$. This identification allows us to see that Eq.~\ref{eq:cssc_l_with_constraint} decribes a superconductor/charged superfluid of charge $Q$. Furthermore, the prefactor is $\sim g^2\sim |\delta|$ gives the superfluid weight, which is, as expected, proportional to the density of the quasiparticles which lead to superfluidity. We may now revisit the term $\frac{k}{4\pi}AdA$ of Eq.~\ref{eq:general_cs_sc_form} in this language, where we simply note that the term $\sim g^2(QA_\mu-\partial_\mu\varphi)^2$ has scaling dimension $\Delta=2$, which is lower than the scaling dimension $\Delta=3$ of the term $AdA$. We can thus expect the superconducting term to be more RG-relevant.

A more powerful argument can be constructed by leveraging the 3D XY particle-vortex duality \cite{PESKIN1978122,Halperin_1981_u1_duality,seiberg_duality_2016,senthil_duality_2019}. This duality establishes that the two Lagrangians $\mathcal L_p$ and $\mathcal{L}_v$ of the form
\begin{equation} \label{eq:particle-vortex}
    \begin{split}
        \mathcal L_p&=|D_A\phi|^2 - r|\phi|^2 + |\phi|^4,\\
        \mathcal L_v&=|D_a\tilde\phi|^2+r|\tilde\phi|^2+|\tilde\phi|^4 + \frac{1}{2\pi}adA,
    \end{split}
\end{equation}
describe the same phase. We focus on the $r>0$ case of this duality. Essentially, a condensed boson $\phi$ charged under $A$ can be described in the dual picture by a gapped boson $\tilde\phi$ which sources vortices and is charged under an internal field $a$. The destruction of order in $\phi$ is linked to the proliferation of vortices sourced by $\tilde\phi$. For our discussion, the mixed Chern-Simons term between $a,A$ on the vortex side is crucial -- imagining $r$ large and positive, we see that $\tilde\phi$ is trivially gapped and may be ignored. We are left with $\mathcal L=\frac{1}{2\pi}adA$, exactly the $Q=1$ case of Eq.~\ref{eq:general_cs_sc_form}, showing that it indeed maps onto a charge-1 superfluid via Eq.~\ref{eq:particle-vortex}. The generalization to other $Q$ follows. The particle-vortex duality involves a mapping of the operators, and the symmetry-broken order parameter $\phi$ on the particle side corresponds to the CS field monopole operator $\mathcal M_a$ on the other side, allowing us to identify the monopole of the internal gauge field $a$ as the order parameter of the superfluid described by Eq.~\ref{eq:general_cs_sc_form}. The mixed Chern-Simons term there dictates that this monopole must have charge $Q$ under the external field $A$ (and it has no charge under $a$), exactly appropriate for a charge-$Q$ order parameter which is gauge invariant under the internal field, as is required for all physical observables.

To conclude this section, we repeat the main point that if a term of the form Eq.~\ref{eq:general_cs_sc_form} appears in a Lagrangian, and if there is no quadratic self-Chern Simons term involving $a$, we have a charged superfluid associated with $a$. The order parameter is the monopole of the $\mathcal{M}_a$ of the gauge field. While Eq.~\ref{eq:general_cs_sc_form} does not encode superfluid weight, we generally expect it to be $\propto\delta$ with $\delta$ the density of doped quasiholes.

\section{Ideal bands with non-uniform Berry curvature} \label{sup:nonuniform_berry}

The aim of this section is to demonstrate that the mean-field doping arguments presented for Landau levels in Sect.~\ref{sub:mfdope} extend more broadly, in particular that they apply to ideal bands. We largely follow \cite{wang_exact_2021,estienne_ideal_2023}. To understand ideal bands, we start with the lowest Landau level with a periodic modulation $V(\mathbf r)$ and one unit of flux per unit cell of $V$. Denoting the generators of this periodicity by $T_{x,y}$, we have $[T_x,V]=[T_y,V]$ with $T_xT_y=T_yT_x$ due to the unit flux. This means that we can diagonalize $T_x,T_y$ simultaneously with $T_x\ket{\psi_\mathbf k}=e^{ik_x}\ket{\psi_\mathbf k}$, $T_y\ket{\psi_\mathbf k}=e^{ik_y}\ket{\psi_\mathbf k}$. These Bloch functions $\ket{\psi_\mathbf k}$ form a Bloch basis for the LLL and may be written in terms of a Jacobi theta function \cite{PhysRevB.31.2529}.

The crucial modification in ideal bands is that the effective field is no longer just $B_0=2\pi/A_\text{u.c.}$, but instead may be taken as $B(\mathbf r)=B_0+\nabla^2K$ where $K(\mathbf r)$ is a periodic function with the same periods as $V(\mathbf r)$. The band is now spanned by $\ket{\varphi_\mathbf k}=e^{-K(\mathbf r)}\ket{\psi_\mathbf k}$. Such functions $\ket{\varphi_\mathbf k}$ are not normalized, with the normalization $\mathcal{N}_\mathbf k$ depending on the momentum, and it is this dependence which leads to the non-uniformity of Berry's curvature \cite{wang_exact_2021}.  While the Berry curvature depends on the unit-cell embedding \cite{Simon_2020}, there exists a preferred embedding in this problem, namely the embedding which makes the band geometry ideal. While one could, in some case,s choose embeddings which make the Berry curvature uniform, the band geometry would then not be ideal in that embedding, making the mapping onto the LLL non-trivial.

But we are interested in a more general problem, in particular in what happens when the field is tuned slightly away from commensurability, $B_0\rightarrow B_0+b$ and $B(\mathbf r)=B_0+b+\nabla^2K(\mathbf r)$. To solve this problem, we may start with a basis of LLL wavefunctions on a torus with $N_\phi$ flux quanta, labelled by $\ket{\psi_\alpha}$ with $\braket{\psi_\beta|\psi_\alpha}=\delta_{\alpha\beta}$. 
In the LLL case, we simply have to solve for the eigenvalues of $H_{\beta\alpha}=\braket{\psi_\beta|V(\mathbf{r})|\psi_\alpha}$ to find the energy levels.

In a non-uniform field, the functions $\ket{\varphi_\alpha}=e^{-K(\mathbf r)}\ket{\psi_\beta}$ span the ideal band, but they are not orthonormal, meaning we must compute the Gram matrix $G_{\beta\alpha}=\braket{\varphi_\beta|\varphi_\alpha}=\braket{\psi_\beta|e^{-2K(\mathbf{r})}|\psi_\alpha}$ and also $V_{\beta\alpha}=\braket{\varphi_\beta|V(\mathbf r)|\varphi_\alpha}=\braket{\psi_\beta|V(\mathbf r)e^{-2K(\mathbf{r})}|\psi_\alpha}$. The physical energy levels are determined by the eigenvalues of $H=G^{-1/2}VG^{-1/2}$.

In the commensurate case of exactly one unit flux per unit cell of $V(\mathbf r)$ and $K(\mathbf r)$, the magnetic eigenstates on a torus (labelled by a Bloch momentum $\mathbf k$) simultaneously diagonalize both $V$ and $G$ with dispersive eigenvalues of $V(\mathbf k)$ and $G(\mathbf k)$, giving a dispersion $\epsilon(\mathbf k)=V(\mathbf k)/G(\mathbf k)$. As in the Landau level case, we focus on a local maximum which we assume isotropic, $\epsilon(\mathbf k_0+\mathbf q)=\epsilon_\text{max.}-\frac{1}{2m^*}|\mathbf q|^2$. The fact that the band maximum is at $\epsilon_\text{max.}$ may be re-stated as the claim that the largest value that $V(\mathbf k)-\epsilon_\text{max.} G(\mathbf k)$ takes on is zero or, equivalently, that the largest eigenvalue of $\tilde V_{\beta\alpha}=\braket{\psi_\beta|[V(\mathbf r)-\epsilon_\text{max.}]e^{-2K(\mathbf r)}|\psi_\alpha}$ is zero. 

When an additional field $b$ is introduced, the eigenvalues of $\tilde V$ behave exactly as the spectrum of a modified potential $\tilde V(\mathbf r)=[V(\mathbf r)-\epsilon_\text{max.}]e^{-2K(\mathbf r)}$ would in a uniform magnetic field. Crucially, because we assumed the spectrum has an isotropic quadratic peak in the commensurate case, this implies that if $b>0$, the peak remains at zero energy. This then proves that the results derived for modulated Landau levels in the main text carry over to the more general ``ideal'' bands. We can expect the mLL spacing pattern of Fig.~\ref{fig:sketch_mlls} to carry over to this case, and so do most of the conclusions above, with the only exception being the fact that the centre of the band, so $\text{Tr}(G^{-1}V)$, moves as $b$ is introduced.

While we derived the above result assuming that the periodic background potential $V(\mathbf r)$ is such that each unit cell encapsulates one flux quantum of the magnetic field, this should really be applied to the parton bands which see flux $1/m$ per unit cell. Taking an $m$-fold increased unit cell, each maximum/minimum is now $m$-fold degenerate. But the above line of reasoning can be applied to each of these extrema individually. Even though the parton unit-cell is of size $m\times1$ and is highly anisotropic, we expect each of the $m$ maxima to be isotropic if the underlying potential $V(\mathbf r)$ has appropriate symmetries.

\section{Flux re-distribution in the mean-field picture} \label{sup:flux_redist}

In Sect.~\ref{sub:mfdope} we used our knowledge of the behaviour of ideal bands in a magnetic field to make predictions for the mean-field behaviour of doped FCI states. One part of the solution was ignored in that analysis --- when constructing a parton mean-field state, one generally has to demand that the densities of the three partons are equal at all points, which is the mean-field approximation to the true constraint that $f_1^\dagger(\mathbf{r})f_1(\mathbf{r})=f_2^\dagger(\mathbf{r})f_2(\mathbf r)=\ldots=f_p^\dagger({\mathbf r})f_p(\mathbf r)$. 

In the states we have predicted to be favourable, the three partons can generally behave differently under doping, which means that this is violated. To restore the constraint at the mean-field level, the effective magnetic field seen by the partons will deviate slightly, $B_p(\mathbf r)=B_0/m +\nabla^2 K/m + b_p + \nabla^2 k_p$ with $k_p$ being chosen to compensate for this density imbalance under the constraint that $\sum_p\nabla^2k_p=0$. A non-zero $k_p$ would then change the mean-field energetics, requiring a self-consistent solution to the problem. While this could be done with an iterative numerical scheme, a simplification occurs for the gapped states where $\tilde\nu_p$ is a multiple of $m$ for all $m$ partons.  The crucial simplification comes from the fact that any density imbalances are suppressed by a factor of $\epsilon=e^{-\pi m/2}$, the origin of which is the fact that the effective magnetic length for the partons is larger than the unit cell by a factor of $\sim\sqrt m$.

Recall that at $\nu=1/m$, the projective translations enforce that each parton to have $m$ degenerate peaks and that the gapped states equally fill the mLL's at each of the peaks. For concreteness, take a square underlying lattice (with lattice constant 1) and take an $m\times1$ extended parton unit-cell. The projective translations mean that the peaks will be at guiding center coordinates $\mathbf R=\mathbf{R}_0+j\hat x$ for $j\in\{0,\ldots,m-1\}$. 

Inserting Landau level orbitals of a magnetic field $2\pi/m$ on such a grid (of unit spacing), it may be checked that the resulting density is $\rho(\mathbf r)=\rho_0+\delta\rho(\mathbf r)$ with $\max|\delta\rho|\sim e^{-m\pi/2}$. While the value of the exponent is a consequence of the square lattice geometry, the same mechanism should hold for other lattice geometries (with the exception of extremely anisotropic limits). This argument is illustrated in Fig.~\ref{fig:m3_uc_sketch}
\begin{figure*}
    \centering
    \includegraphics[width=0.94\linewidth]{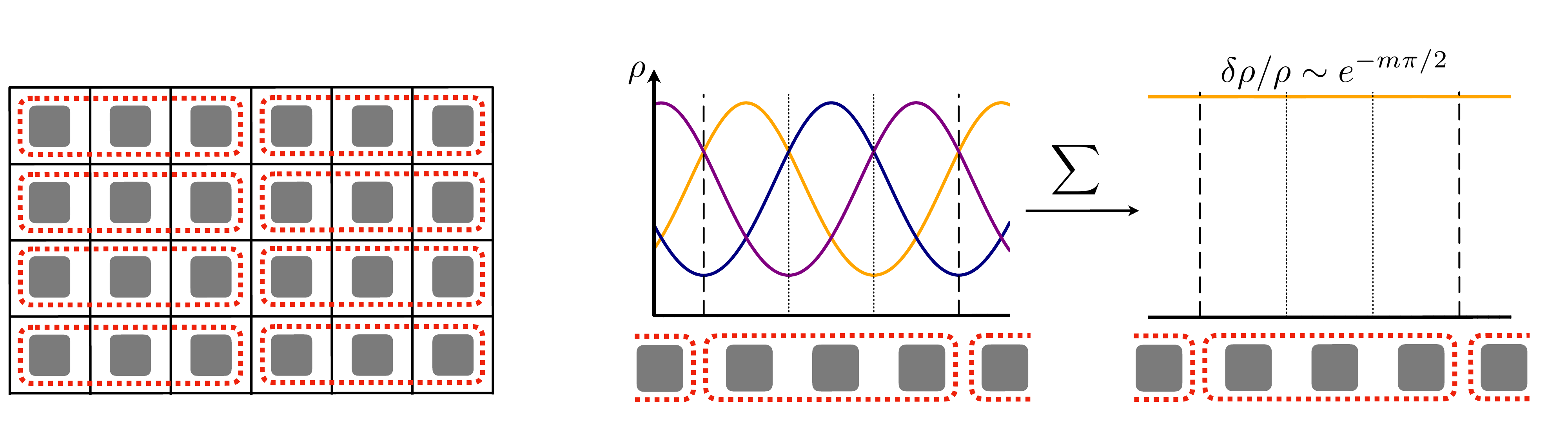}
    \caption{Real-space density in a parton band at $m=3$. Left: taking an original unit cell generated by $T_x,T_y$, the projective translation symmetry $T_xT_yT_x^{-1}T_y^{-1}=e^{2\pi i/3}$ forces us to consider dispersion within a larger parton unit cell generated by $T_x^3,T_y$. But because $T_x$ is still a symmetry, this gives a 3-fold degeneracy of the orbitals. Right: We plot the real-space density of three such orbitals related by $T_x$ in the case of the LLL. In gapped translationally invariant states, all these orbitals are filled, and the resulting density is the sum of densities of these orbitals, which is close to constant, with fluctuations $\delta\rho/\rho\sim\epsilon=e^{-m\pi/2}$ which are invariant under $T_x,T_y$, unlike the densities of the single parton orbitals.   
    }
    \label{fig:m3_uc_sketch}
\end{figure*}

Combining with the density $\delta$ of holes gives an expectation that the difference in parton densities if all $f_p\rightarrow0$ will be $|n_p(\mathbf r)-\bar n(\mathbf r)|\sim \delta\epsilon$, which suggests that $|k_p|\sim \delta\epsilon$ to compensate. The change in effective potential felt by the partons is then $\sim V_0\delta\epsilon^2$ with another factor of $\epsilon$ coming from projecting the real-space potential into the lowest Landau level. With $\sim\delta$ holes in each parton band, we get an estimate $\Delta E_\text{redist.}\sim\epsilon^2\delta^2 V_0$ for the change in energy due to flux re-distribution effects.

For comparison, the background potential at scale $V_0$ gives an energy scale $E_0\sim\epsilon V_0$ upon projecting into the lowest parton Landau level and the discussion in the main text shows that the different states have energy relative differences $\propto\delta^2$, so $E_0\sim \epsilon\delta^2V_0$, giving us $\Delta E_\text{redist.}\sim \epsilon E_0$ justifynig that this re-distribution may be ignored at lowest order. For completeness, we note that a non-uniform background field (with the relative scale of modulations $\kappa$) leads to an energy change $\sim V_0\kappa \delta^2\epsilon\sim \kappa E_0$ with the first three factors coming from the discussion around Eq.~\ref{eq:modification_nonuniform_field} and the final factor of $\epsilon$ resulting from the kernel (with characteristic lengthscale $\sim\sqrt m$) present in the integrals. Thus, even for a close-to-uniform background field at small $\kappa$, these effects dominate over the flux-redistribution scale $\Delta E_\text{redist.}$.

\section{Variational Monte Carlo details} \label{sup:vmc}
We work on a lattice model whose lowest band is an ideal $\chern=1$ band, closely resembling a generalized Landau level. To construct such models in practice, we follow Kapit and Mueller \cite{kapit_exact_2010} --- they consider a 2D grid of points with the coordinates written as $z_j=u_j+iv_j$ where $(u_j,v_j)\in\mathbb Z^2$ specify the site. The idea is to design the hoppings such that the wavefunctions of the lowest bands are exactly the same as the LLL wavefunctions in some magnetic field $2\pi\phi$, evaluated on the $\{z_j\}$. This may be done with a Hamiltonian of the form $H=\sum_{ij}H^\text{KM}_{ij}c_i^\dagger c_j$. It was shown that if one takes (with $z=xd_x+iyd_y$)
\begin{equation} \label{eq:kmhamwrittenout}
\begin{split}
    H^\text{KM}_{ij}&=W(z_i-z_j)e^{(\pi/2)(z_jz_k^*-z_j^*z_k)\phi}\\
    W(z)&=e^{-\frac\pi2\left[(1-\phi)|z|^2\right]}(-1)^{\text{Re} z +\text{Im}z+\text{Re}z\text{Im}z},
\end{split}
\end{equation}
there exists an exactly flat band at zero energy whose wavefunctions are $\psi^{(k)}_i=\psi^{(k)}(z_i)=z_i^k e^{-\pi\phi|z_i|^2/2}$, exactly the appropriate LLL. Recall that $x,y$ coordinates of the sites are spaced by $(1/d_x,1/d_y)$ in our construction, meaning that the real and imaginary parts of $z$ are both always integers [and that for any $(u,v)\in\mathbb Z^2$, there exists a site at $(x,y)=(u/d_x,v/d_y)$]. The band being flat and at zero energy means that $H_{ij}^\text{KM} \psi^{(k)}_j = 0$ (with summing over $j$ implied). Of course, Eq.~\ref{eq:kmhamwrittenout} includes hopping terms over arbitrary distances --- but note that due to the Gaussian decay of $W(z)$, the maximal hopping range may be truncated without introducing a significant error. In our numerics, $\phi=1/d_xd_y=1/9,$ and we keep hopping terms up to $|z|^2=4$, as that already creates a sufficiently flat band. In a generalized Landau level, we want the wavefunctions to instead be of the form $\tilde\psi^{(k)}_i=e^{-K(\mathbf r_i)}\psi_i^{(k)}$. This may be achieved by a Hamiltonian $\tilde H_{ij}=e^{K(\mathbf r_i)}H^\text{KM}_{ij}e^{K(\mathbf r_j)}$, where it is clear to see that now $\tilde H_{ij}\tilde\psi_j^{(k)}=0$, so we still have a flat band which is spanned by a generalized Landau level. If $K(\mathbf r)$ is lattice periodic, as is the case in our work (and if there is one quantum of flux per unit cell in the mapped Landau level), the arguments of \cite{wang_exact_2021,estienne_ideal_2023} show that we have constructed a perfectly flat ideal $\chern=1$ band.

With a microscopic lattice model in mind, we can discuss how the VMC optimization works. We closely follow the procedure outlined in the Supplement of \cite{lotric_paired_2025}. Given some number of partons and arrangement of mean-field fluxes, we first find hopping phases $e^{i\phi_{ij}^{(p)}}$ for parton $p$ moving between sites $i,j$ such that the effective field felt by this parton is exactly the required mean-field flux. In particular, we want to enforce a flux density $\phi_p=2\pi/m+b_p$ on each parton, and this may be achieved by letting
\begin{equation}
    \phi_{ij}^{(p)} = (2\pi/m+b_p) (z_i z_j^* - z_i^*z_j)/4,
\end{equation}
or by any other assignment related to this by a gauge transformation $\phi^{(p)}_{ij}\rightarrow\phi^{(p)}_{ij}+\lambda_i-\lambda_j$. We then initialize a mean-field Hamiltonian $\tilde H_\text{MF}^{(p)}$ for each parton such that $[\tilde H_\text{MF}^{(p)},T_{x,y}]=0$, i.e.~a Hamiltonian translationally invariant with respect to the translation group of $\tilde H_{ij}$. This guarantees that $H^{(p)}_{\text{MF};ij}=\tilde H^{(p)}_{\text{MF};ij}e^{i\phi^{(p)}_{ij}}$ describes a Hamiltonian with the correct mean-field flux and that it obeys the \textit{projective} translation symmetry $[H^{(p)}_\text{MF},~T_{x/y}^{(p)}]=0$ with projective translation operators $T_x^{(p)}T_y^{(p)}=e^{i\phi_p}T_y^{(p)}T_x^{(p)}$. The translationally invariant hopping parameters (up to a limited range) in $\tilde H^{(p)}_{\text{MF};ij}$ are taken to be variational parameters for our state.  With $N$ particles, we simply take the $N$ lowest-energy eigenvectors of $\tilde H^{(p)}$ for our parton mean-field state (depending on the parton's observed flux, this may or may not be a full band) and use Eq.~\ref{eq:partonform} to construct the microscopic wavefunction.  The optimization procedure evaluates gradients of the energy with respect to the parton eigenvectors, from which we can, in turn, infer how $H^{(p)}_\text{MF}$ should be adjusted. We repeat this procedure until convergence, remembering to impose translational invariance of the parton Hamiltonian without the additional gauge field at every step \cite{lotric_paired_2025}. This procedure should yield the lowest-energy parton state in each mean-field flux sector.

The energy of the fermionic ansatz can be improved by adding a Jastrow factor, $\psi(\mathbf r_1,\mathbf r_2,\ldots)\rightarrow\prod_{i<j}e^{-J(\mathbf r_i,\mathbf r_j)}\psi(\mathbf r_1,\mathbf r_2,\ldots)$ with the Jastrow factor $J(\mathbf r,\mathbf r')$ a real function. We demand that $J$ is translationally invariant under $T_{x,y}$ and that it is local, $J(\mathbf r,\mathbf r')=0$ if $|\mathbf r - \mathbf r'|>l_\text{Jastrow}$. Outside of these constraints, the values of $J$ are also optimized over. In our numerical calculations, we set $l_\text{Jastrow}=1$ in units where the nearest neighbouring sites are separated by a distance $1/d_x=1/d_y=1/3$.

The optimization is carried out via stochastic gradient descent --- we may generally write our variational ansatz as $\psi_\theta(\{\mathbf r_i\})$ where $\theta$ encodes the variational parameters. We carry out Markov Chain Monte Carlo sampling according to the distribution $p_\theta(\{\mathbf r_i\})=|\psi_\theta(\{\mathbf r_i\})|^2$ and at each sample compute the local energy $E_L=\braket{\{\mathbf r_i\}|H|\psi_\theta}/\braket{\{\mathbf r_i\}|\psi_\theta}$ and the gradients of the wavefunction with respect to the variational parameters $\mathcal{O}_l=\frac{\partial}{\partial\theta_l}\log \psi_\theta$. From these quantities, we can compute the energy of the state $E_\theta=\langle E_L\rangle$, while the ``forces'' $F_l=\langle E_L\mathcal{O}_l^*\rangle-\langle E_L\rangle\langle\mathcal{O}_l^*\rangle$ determine the derivatives of $E_\theta$ with respect to the variational parameters. We use the ADAM optimizer \cite{37_kingma2017adammethodstochasticoptimization}, which updates $\theta_l$ based on a rolling average of the $F_l$ over multiple learning iterations. Operationally, for each model, we begin by optimizing $\theta$ for $\nu=\frac1m$ and use that to initialize the different runs at $\nu=\frac1m-\delta$ in order to save computational effort.

}

\bibliography{references}

\include{supplement}

\end{document}

%% file: supplement.tex
\onecolumngrid
\newpage
\clearpage

\begin{center}
    {\large\bf Supplementary Material for:}

\vspace*{10pt}

{\large \bf Phases of itinerant anyons in Laughlin's quantum Hall states on a lattice}

\begin{center}
   Tevž Lotrič and Steven H. Simon 
\end{center}

\vspace*{10pt}
    
\end{center}

\setcounter{section}{0}
\setcounter{equation}{0}
\setcounter{figure}{0}
\setcounter{table}{0}
\setcounter{page}{1}
\makeatletter
\renewcommand{\theequation}{S\arabic{equation}}
\renewcommand{\thefigure}{S\arabic{figure}}
\renewcommand{\thetable}{S\arabic{table}}

\section{Recovering the full low-energy state manifold with localized quasiholes states} \label{sub:local_QHs}
The states we have considered so far all share the property that the quasiholes are not the familiar quantum Hall localized anyons, but instead are delocalized objects where the flux associated with them is not directly pinned to the partons. In this section, we consider the more typical localized anyon \textit{Ansätze}, working in the $K_0=0.25$ case of our generalized Kapit-Mueller model discussed in Sect.~\ref{sect:vmc} of the main text and Appendix~\ref{sup:vmc}. This section aims to build confidence in the parton approach, which we do by demonstrating that we are able to accurately construct the entire low-energy subspace of states with 2 or 3 quasiholes. 

\subsection{Counting of states} \label{supsub:count}

We start by reviewing the analytical expectations. Consider an FCI on a torus with $N_\phi$ unit cells (or a Landau Level with $N_\phi$ flux quanta) and $N$ bosons. When $N_\phi=2N$, we have the undoped $\nu=\frac12$ FCI and here, we consider cases with $N_\phi-2N=Q>0$ quasiholes. How many low-energy states do we expect? By low-energy, we mean below the many-body gap $\Delta$ relative to the ground state --- in the FQH case with the model interaction, these states are all degenerate, but in an FCI, they will split by an energy related to the anyon dispersion, denoted by $\sim V$ in the rest of the paper. The counting is most transparent when we take the FQH thin-torus limit\cite{Regnault_2011}, which turns the FCI into a period-2 charge-density wave. The occupation pattern of the ground state is $\cdots 01010101\cdots$ and the number of low-energy states is the solution to the counting problem of how many ways there are to occupy $N$ of the $N_\phi$ orbitals arranged on a circle in such a way that no two neighbouring orbitals are both occupied. 

To insert $Q$ anyons, we start by giving them labels $1\ldots Q$ and start by placing anyon ``1'' on any of the $N_\phi$ sites (in the occupation pattern, an anyon can be seen as a double zero, $\cdots0101\underline{0}0101$, where the underlined orbital is taken as the anyon's location. The next anyon, ``2'' will be placed at a different position, with $0,\dots,N$ bosons between the two -- this gives a factor $\times (N+1)$. For anyon ``3'', we again choose between $0,\dots,N$ anyons separating it from ``1'', but if it happens to fall in the same gap as ``2'', we have to specify their ordering, as our anyons are labelled, giving $(N+2)$ options for anyon ``3'' and for each following anyon $q$, we get $(N-1+q)$ further options. Finally, dividing out the $Q!$ anyon permutations, we arrive at the number of states in the low-energy manifold for $Q$ anyons and $N$ bosons (in $N_\phi=2N+Q$ flux) as
\begin{equation} \label{eq:lowenergystatecount}
    D(Q,N)=(2N+Q)\times(N+1)\times(N+2)\times\ldots\times(N+Q-1)/Q!=\binom{N+Q}{Q}+\binom{N+Q-1}{Q}.
\end{equation}
The cases we check numerically are $D(2,N)=(N+1)^2$ and $D(3,N)=(2N+3)(N+1)(N+2)/6$. In the FQHE problem, we know how to construct states with localized quasiholes, using the wavefunction $\chi^{uv}$ in Eq.~\ref{eq:two_local_holes} which constructs holes at $u,v$. The set of wavefunctions $\{\chi^{u_0,v_0},\chi^{u_1,v_1},\ldots, \chi^{u_T,v_T}\}$ with holes at different positions $(u_j,v_j)$, will, for sufficiently large $T>D(2,N)$ (and sufficiently varied positions) form an over-complete basis of two-quasiholes states. We may then compute the Gram matrix $G_{ij}=\braket{\chi^{u_i,v_i}|\chi^{u_j,v_j}}$ and the number of non-zero eigenvalues of $G$ tells us the dimension of the low-energy manifold.

\subsection{Construction of local-quasihole states} \label{supsub:constr}

To carry out this procedure with the parton construction in the FCI case, we must first understand how to construct local-quasihole states in the parton construction. A quasihole at position $u$ corresponds to multiplying the Laughlin wavefunction by $\prod_i(z_i-u)$. One may attempt to construct a parton state by associating this factor to one of the partons, so we would have $\Omega_1=\prod_i(z_i-u)\prod_{i<j}(z_i-z_j)$ and $\Omega_2=\prod_{i<j}(z_i-z_j)$, but unlike in Sect.~\ref{sup:interactions_effect} of the Supplement, here we do not delocalize the quasihole, as there is not an integral over $u$. This then implies that the density of the two partons is different -- $n_1(z)\rightarrow |z-u|^2$ as $z\rightarrow u$, while $n_2$ remains constant. This goes against the constraint $f_1^\dagger f_1=f_2^\dagger f_2$, which we hope to enforce at least in an averaged sense. 

To fix this discrepancy, we can insert a half flux quantum $\pi$ of the internal gauge field at the point $u$, which introduces factors $\prod_{i}(z_i-u)^{-\frac12}$ and $\prod_i(z_i-u)^{\frac12}$ to the parton wavefunctions. Combining with a particle removed for parton 1 at $u$, we find $\tilde\Omega_1=\tilde\Omega_2=\prod_i\sqrt{z_i-u} \prod_{i<j}(z_i-z_j)$, where the product of the two correctly reproduces a Laughlin quasihole, and the two partons have equal density everywhere. There are branch cuts in the parton wavefunctions which reflect the semionic statistics, but the full \textit{Ansatz} for the wavefunctions is well-defined, as the branch cuts cancel. In practice, we may pick two locations $u,v$ and insert $\pi$-fluxes there, resulting in $N$ particles for each parton, with parton 1 seeing $N+2$ flux and parton 2 seeing $N$ flux. We have two holes available in parton 1, and we can expect that when solving self-consistently, they will be attracted to the $\pi$-fluxes to cancel the $\prod_i(z_i-u)^{-\frac12}$ factors which would be present otherwise. So in our solver, we simply impose $\pi$-fluxes on some locations and optimize the parton \textit{Ansatz}, which will in turn remove partons from the appropriate locations.

Another argument for why this is the correct construction can be made in the thin-torus limit, where the occupation pattern around an isolated quasihole is $\cdots \underline{01}~\underline{01}~\underline{00}~\underline{10}~\underline{10}\cdots$. Each parton effectively sees half flux, so for each underlined pair of orbitals in this thin-torus limit, the parton only has one orbital. It is then clear to see that a quasihole is the combination of an empty parton orbital \textit{and} a half-period shift for the parton orbitals to one side, consistent with a $\pi$-flux inserted at the quasihole.

To summarize, Sect.~\ref{supsub:count} provides a prediction for the number of states in the low-energy manifold of a FQH state (the FCI result is expected to be equal) while in Sect.~\ref{supsub:constr} we have argued that to construct these states, we should demand localized $\pi$-fluxes at some positions, and the resulting state will have quasiholes localized there.  The basis presented here exhausts the set of low-energy eigenstates of the system, which means that any of the itinerant anyon phases discussed in the main text may be written as a superposition of these states --- in this way, the discussion in this paper may be re-phrased as asking what form of quasihole wavefunction superposition is favourable, as we briefly touch on in Sect.~\ref{sup:interactions_effect} of the Supplement as well.

\subsection{Numerical results}
We finally show the numerical results. Details of how this is done are exactly as in Ref.~\cite{lotric_paired_2025} and as is reviewed in Appendix~\ref{sup:vmc} above. Note that to make the enforcement of the parton $\pi$-fluxes transparent, we keep the parton mean-field Hamiltonian (before flux insertion) $\tilde H_{\text{MF};ij}^{(p)}$ translationally invariant, even though it strictly does not have to be (and we expect that the quality of the state can be improved by relaxing this constraint). The justification is that the results we get this way are already good enough to demonstrate that the parton approach works well in this case.

\begin{figure}
    \centering
    \includegraphics[width=0.9\linewidth]{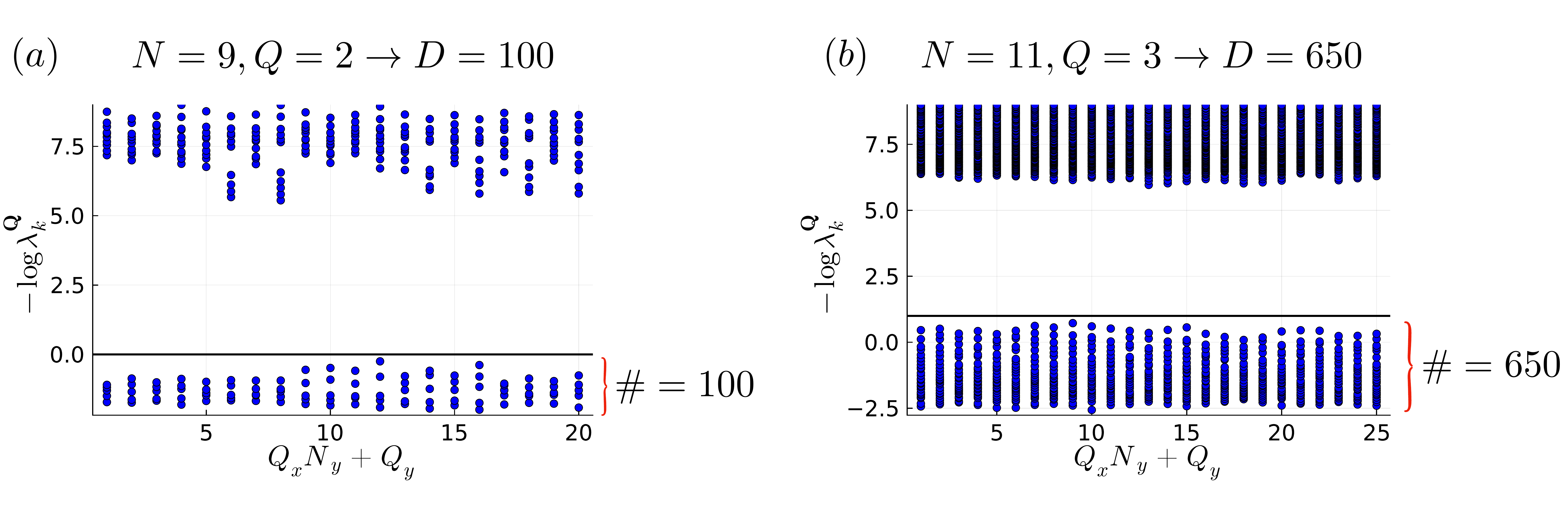}
    \caption{Momentum-resolved eigenvalues of the Gram matrices for localized-quasihole states on the generalized Kapit-Mueller model (with $K_0=0.25$) for (a) a 2-quasihole state (on a $5\times4$ torus) and (b) a 3-quasihole FCI (on a $5\times5$ torus). The results indicate a large gap in the spectrum, suggesting that only the large eigenvalues below the gap in $\log\lambda^\mathbf Q_k$ (below the black line in the plot) correspond to the physical low-energy states, while the remaining small, but non-zero eigenvalues above the line are due to the fact that the \textit{Ansatz} does not capture the low-energy space exactly and also due to Monte-Carlo sampling. In both cases, the number of physical eigenstates exactly matches the expected dimensionality $D(Q,N)$ (Eq.~\ref{eq:lowenergystatecount}) of the low-energy Hilbert space -- in case (b), our method remarkably correctly predicts 650 orthogonal states. 
    }
    \label{fig:grammatrix}
\end{figure}

Given $Q$ quasiholes, the $\pi$-fluxes are inserted at unit cells $\{(x_1,y_1),\dots,(x_Q,y_Q)\}$. For each such configuration, we do a VMC run to optimize the effective parton mean-field Hamiltonian. Varying over all physically distinct configurations (which are inequivalent under permutations of the quasiholes and under centre-of-mass translations) constructs a set of states $\{\ket \psi_i\}$, and we decompose each of these states $\ket{\psi_i}$ into components of different values of the centre-of-mass momentum $\mathbf Q$, labelled by $\ket{\psi_i^\mathbf Q}$ -- even if the state is uniform, $\mathbf Q$ is a good quantum number due to the translation invariant FCI Hamiltonian. At each $\mathbf Q$, we compute the Gram matrix $G_{ij}^\mathbf Q=\braket{\psi_i^\mathbf Q|\psi_j^\mathbf Q}$. We take the eigenvalues of $G_{ij}^\mathbf Q$, labelled by $\lambda^\mathbf Q_k$ and plot $\log\lambda^\mathbf Q_k$ versus $\mathbf Q$ in Fig.~\ref{fig:grammatrix}. 
The number of non-zero eigenvalues tells us the number of orthogonal states we can construct, and the results in Fig.~\ref{fig:grammatrix} clearly indicate excellent agreement with the expectation.

\begin{figure}
    \centering
    \includegraphics[width=0.9\linewidth]{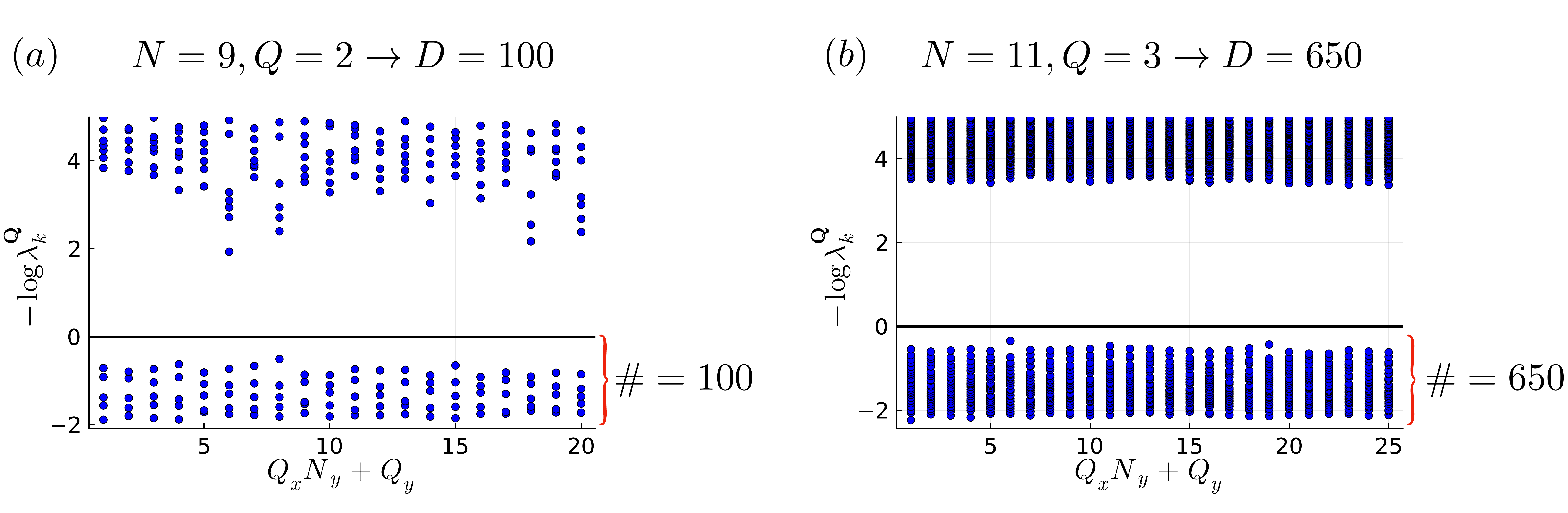}
    \caption{Momentum-resolved eigenvalues of the Gram matrices for localized-quasihole states on the ``checkerboard'' model for (a) a 2-quasihole state (on a $5\times4$ torus) and (b) a 3-quasihole FCI (on a $5\times5$ torus). The analysis is identical to Fig.~\ref{fig:grammatrix}, with the only difference being in which microscopic model we consider. The results again a large gap in the spectrum, suggesting that only the large eigenvalues below the $\log\lambda^\mathbf Q_k=0$ line are in the physical low-energy states, while the remaining small, but non-zero eigenvalues above the line are not physical states. We again see exact agreement with the expected dimensionality $D(Q,N)$ (Eq.~\ref{eq:lowenergystatecount}) of the low-energy Hilbert space. For this model, we compute the overlap with the true wavefunctions of the low-energy subspace (obtained via exact diagonalization) and find on average a $96\%$ overlap, showing that our procedure not only constructs the correct number of states, but that it also constructs all those states to high precision.
    }
    \label{fig:grammatrix_check}
\end{figure}

In our previous related work, Ref.~\cite{lotric_paired_2025}, we studied a ``checkerboard'' model of an FCI -- while we do not review this model here, the important feature is that it is a two-band model, where the lower $\chern=1$ band is close to flat, and in this model, numerical calculations have confirmed a bosonic $\nu=\frac12$ FCI, despite the fact that this lower band is not perfectly flat and that it does not posses an ideal or vortexable geometry. Still, in Ref.~\cite{lotric_paired_2025} we have found the parton description to work remarkably well for a range of parameters in the undoped model. Here we extend this analysis to local-quasihole states -- in Fig.~\ref{fig:grammatrix_check}, we repeat the analysis we did for the Kapit-Mueller model in Fig.~\ref{fig:grammatrix} for this checkerboard model.
For the checkerboard model on such small systems, we may take the predicted physical states (the ones corresponding to large $\lambda^\mathbf Q_k$), labelled by $\ket{\phi^\mathbf Q_k}$ and compute their overlap with the exact low-energy states at the given momentum, $\ket{v^\mathbf Q_l}$, obtained using the readily available DiagHam library. In particular, defining $O_k^\mathbf Q=\sum_l |\braket{v^\mathbf Q_l|\phi^\mathbf Q_k}|^2$ as the overlap of the state $(\mathbf Q,k)$ with the exact low-energy state manifold. We find that for the physical states in Fig.~\ref{fig:grammatrix}, the average value of $O^\mathbf Q_k$ is $96\%$, indicating that we not only recover the correct number of orthogonal states, but that these states are also in good agreement with the exact low-energy states.  If we relaxed the translational invariance constraint on $\tilde H_{\text {MF};ij}^{(p)}$, the additional variational parameters should allow for a lower variational energy and thus an even larger overlap with the exact low-energy states. 
This good agreement with exact diagonalization for the dispersive, non-ideal checkerboard model, for which finding good trial wavefunctions can be very tricky, is highly encouraging. While we do not perform exact diagonalization calculations on the generalized Kapit-Mueller model, the combination of this high overlap for the checkerboard model and the fact that our construction is exact in the limit of a Landau Level should give confidence that our results for this generalized KM model are correct to high accuracy, as this model shares more features with the LL than the checkerboard does (in particular, the generalized KM model has an ideal band geometry).

\section{Chern-Simons response theory for gapped states at $\nu=\frac13-\delta$} \label{sup:cs_general_sc}
In the main text, we saw that when doping the $\nu=\frac13$ state such that the quasiholes are at fillings $\tilde\nu=(-3,-3,0)$, the resulting state is a superconductor. In this section, we demonstrate that whenever all $\tilde\nu$ are integers, the same conclusion results.  The projective translation symmetry for the partons implies three degenerate maxima in each parton's dispersion, meaning that if we demand that an integer number of mLL's is filled in each valley (to achieve a state which is gapped and does not break translational symmetry), all $\tilde\nu$ must be multiples of three. But the superconducting conclusion arises more generally, even if they are not, for example, by the translation symmetry being broken, which could lead to a unique maximum for the parton dispersions. So all $\tilde\nu$ being multiples of three guarantees a gapped mean-field \textit{Ansatz}, and (independently) a gapped \textit{Ansatz} and all $\tilde\nu$ being integers leads to a superconducting state. So together, $\tilde\nu$ all being multiples of three on its own guarantees superconductivity, while if $\tilde\nu$ are all integers, superconductivity is a possibility if the parton mean-field can become gapped, for example via a spontaneous breaking of the translational symmetry.

Upon doping a density $\delta$ of holes, denote the fluxes seen by the partons with $b_p$ with the constraint $\sum_pb_p=0$. Each parton on average sees $\delta+b_p/2\pi$ quasiholes in flux $-b_p$, leading to a filling fraction $\tilde\nu_p=-2\pi\delta/b_p-1$. The gapped states happen when all $b_p\neq0$, so we may set $(L,M,N)=(-2\pi\delta/b_1,-2\pi\delta/b_2,-2\pi\delta/b_3)$ where we require $L,M,N$ integer (to ensure integer hole fillings) and also that $\sum_pb_p=0$, which leads to $LM+MN+NL=0$. The partons fill $\mathcal{C}=1$ bands with holes at fillings $-1+(L,M,N)$ on top, meaning that Chern number of the gapped partons are $\mathcal{C}=(L,M,N)$ and the gaps are $\Delta\sim(1/|L|,1/|M|,1/|N|)$. Assuming a $\U(1)\times \U(1)$ invariant gauge group for now, we may write the effective theory, assigning partons to be charged under $(a_1,a_2-a_1,A-a_2)$
\begin{equation}
\begin{split}
    \mathcal{L}&=\frac{1}{4\pi}a_I K_{IJ}da_J - \frac{1}{2\pi}t_Ia_IdA + \frac{N}{4\pi}AdA,\\
    K&=\begin{pmatrix}
       L+M & -M \\
       -M & M+N
    \end{pmatrix}\hspace{20px}t=(0,N).
\end{split}
\end{equation}
Because $\det K=0$, there exists a zero eigenvalue and the combination of gauge fields associated with that eigenvalue will lead to superconductivity. We may parametrize all solutions to the Diophantine equation $LM+MN+NL=0$ as $k,u,v\in\mathbb{Z}$ with $\gcd(u,v)=1$ and
$L=k(u+v)u,~M=-kuv,~N=k(u+v)v$. Applying the transformation $W=\begin{pmatrix}
    v &-u\\
    -\rho & \sigma
\end{pmatrix}\in SL(\mathbb Z,2)$ with $v\sigma=1+u\rho$ (such $\sigma,\rho\in\mathbb Z$ always exist if $\gcd(u,v)=1$) by letting $K\rightarrow WKW^T $, $t\rightarrow Wt$ and $a\rightarrow Wa$, we find that with $\tilde a_1=va_1-ua_2$ and $\tilde a_2=\sigma a_2-\rho a_1$,
\begin{equation}
    \mathcal{L}=\frac{k}{4\pi}\tilde a_2 d\tilde a_2 - \frac{kv(u+v)\sigma}{2\pi}\tilde a_2 dA + \frac{Q}{2\pi}\tilde a_1 dA
\end{equation}
Because there is no quadratic term for $\tilde a_1$, this Higgses the external electromagnetic field $A$ and leads to a superconductor. The order parameter is the monopole operator $\mathcal M_{\tilde a_1}$ and it has charge $Q=kuv(u+v)$, which is re-assuringly always an even number. While a superconductor exists for any $k,u,v\in\mathbb Z$ if $\gcd(u,v)=1$, we can expect the most important cases to be ones where the gap is larger, which translates to small $|L|,|M|,|N|$, meaning that smaller $k,u,v$ generally give lower-energy states.

One case is $u=-2,v=1,k=-1$ which gives $\tilde\nu=(-3,-3,0)$ and is the translationally invariant superconductor discussed in the main text (up to parton permutations). Because two of the partons have identical $\tilde\nu$ and arise from an identical underlying band, they may have an identical mean-field \textit{Ansätze} which would in turn imply a $\U(2)$ gauge invariance. If there is $\U(2)$ gauge invariance in the $\tilde\nu=(-3,-3,0)$ state, the resulting central charge is $c_-=\frac52$, and the state is related to one of the constructions in Ref.~\cite{shi_non-abelian_2025} of states achieved by tuning bandwidth. We note that $\U(2)$ is the largest gauge group that these gapped states can have when starting from $m=3$ --- the cases where the larger $\SU(3)$ remains unbroken are not gapped at the mean field level as discussed in Sect.~\ref{sect:undistr} in the main text.

Another state worth pointing out is $u=-2,~v=1,~k=1$ with $\tilde\nu=(1,1,-2)$. Because $f_1,f_2$ have the same Chern number (and both arise from the same underlying band, doped in the same way), they might again have an identical mean-field \textit{Ansatz}, leading to $\U(2)$ gauge invariance with the $\SU(2)$ component acting on $(f_1,f_2)$. We note that this state has  $\mathcal{C}=(2,2,-1)$, so the response is the time-reversal conjugate of the $\tilde\nu=(-3,-3,0)$ state which has $\mathcal C = (-2,-2,1)$. From this, we expect $c_-=+3/2$ for the $\tilde\nu=(1,1,-2)$ state, which, after particle-hole conjugating the entire Landau level, gives a doped $\nu=\frac23+\delta$ state which is a charge-2 superconductor with central charge $\tilde c_-=1-\frac32=-\frac12$. This agrees with the value inferred based on pairing symmetry in Ref.~\cite{wang_chiral_2025}. But since $\tilde\nu$ are not multiples of $m=3$ in this state, it likely breaks translational symmetry. But despite this, we argue that the effects of this symmetry breaking may be relatively mild, in particular that they may only be of $\mathcal{O}(\delta^2)$ for a density $\delta$ of quasiholes. To understand this, note that the $\tilde\nu=(1,1,-2)$ \textit{Ansatz} has quasihole densities $\delta/2$, $\delta/2$ and $2\delta$. Out of the three degenerate peaks available in the dispersion for each parton (see Fig.~\ref{fig:m3_uc_sketch}), put all the quasiholes arising from partons 1,2 onto the first (orange) peak while splitting parton 3's quasiholes among the other two peaks. While such a construction only respects $T_x^3$ for any given parton, summing over the partons cancels these deviations to lowest order (as each peak has a density $\delta$ of holes), while the different effective fields seen by the partons mean that the profiles do not cancel perfectly at $\mathcal{O}(\delta^2)$. While our numerics have not identified scenarios where this state would be favourable, it is worth noting its existence due to the DMRG results of \cite{wang_chiral_2025}.

\section{Anyon interpretation of the superconductor from $\nu=\frac13-\delta$} \label{sup:stack_and_condense}

The $\U(2)$ anyon superconducting state we found when hole-doping the $\nu=\frac13$ FCI at $\tilde\nu=(-3,-3,0)$ analytically in Sect.~\ref{sub:mfdope} and within VMC in Sect.~\ref{sect:vmc} deserves more discussion. Its existence is rather surprising in light of the expectation that since three $Q=\frac13$ anyons can form a microscopic fermion, there is no reason to combine further anyons into Cooper pairs. We attempt an explanation of this in the ``stack and condense'' picture developed in Ref.~\cite{senthil_doping_2025,seo_unified_2026}. We first give the rough outline of the logic, followed by a more detailed calculation in Sect.~\ref{supsub:stack_condense_details}. We recall that the starting point is the Laughlin state which may be described by $\U(2)_{-1,-6}$ Chern-Simons theory, {where writing $\U(2)=[\SU(2)\times \U(1)]/\mathbb Z_2$, we can follow \cite{senthil_doping_2025,shi_non-abelian_2025} to define $\U(2)_{k_1,k_2}=[\SU(2)_{k_1}\times \U(1)_{k_2}]/\mathbb Z_2$, which is a valid theory if $2k_1+k_2\in 4\mathbb Z$ [it is bosonic (fermionic) if $(2k_1+k_2)/4$ is even (odd)]}. This $\U(2)_{-1,-6}$ theory has the same anyon content as the more usual $\U(1)_3$ picture. Our construction then ``stacks'' a neutral $\U(2)_{3,6}$ topological order created by the quasiholes on top of this. This order is generated by three anyon types, $h,c,\tau$, with fusion rules $h^3=I,~c^2=I,\tau^2=I+\tau$. At finite anyon density, the degeneracy of the two fusion channels is lifted and we assume the ``paramagnetic'' $I$ is favoured over $\tau$ \cite{senthil_doping_2025,seo_unified_2026} (the contrary leads to a spontaneous breaking $\U(2)\rightarrow \U(1)\times \U(1)$ and ultimately an imbalance of the mean-field flux between $f_1,f_2$). Denoting the charge $e/3$ anyon in $\U(2)_{-1,-6}$ by $q$, we argue that the elementary charged excitation of the stacked theory is $\bar qv$ with $v=hc\tau$. Physically, we may picture that each $Q=\frac13$ quasiparticle introduced into the system in the  $\U(2)_{-1,-6}$ order splits into (1) a charged part and (2) the neutral anyon $v=hc\tau$. Combining three quasiholes gives a unit charge and $v^3\rightarrow c\tau$ which is a non-trivial anyon (assuming $\tau\times\tau\rightarrow I$). To be able to fuse into the identity, we require at least $v^6$, meaning that one needs to combine 6 original Laughlin quasiholes to create an energetically favourable local excitation in the stacked order, implying that the lowest-charge local excitation in the low-energy theory is in fact the charge-$2e$ Cooper pair. 

While this could be part of the explanation, we note that it is presently not clear how this translates to the algebraic theory of anyon condensation --- while a clear correspondence between anyon superconductivity and anyon condensation exists for Abelian CS theory, the mapping in the non-Abelian case appears more complicated \cite{shi_non-abelian_2025}. With the above picture outlined, we now turn to a more detailed calculation.

\subsection{Details of the stack-and-condense calculation} \label{supsub:stack_condense_details}
We follow \cite{senthil_doping_2025,seo_unified_2026} closely, starting with the undoped $\nu=\frac13$ FCI which we choose to describe by a $\U(2)$ gauge field $\alpha$ -- note that in contrast to Eq.~\ref{eq:one_third_u2_sc_state} of the main text, the gauge field $\alpha$ here is $\U(2)$ and not $\SU(2)$, meaning that we allow for $\text{Tr}\alpha\neq0$ instead of introducing an additional field $a$, the reason being that this unified picture, while not separating the charge and neutral sectors as neatly, correctly accounts for the $\mathbb Z_2$ quotient in $\U(2)=[\U(1)\times \SU(2)]/\mathbb Z_2$, which is a detail we mostly ignored in the main text, as the conclusions drawn there were independent of it. The partons $(f_1,f_2)$ form the $(j,n)=(1/2,1)$ representation of $\U(2)$ while $f_3$ is in the $(0,-2)$, making $\psi=f_1f_2f_3$ a singlet as needed. In the undoped case, all the partons are in gapped bands with $\mathcal{C}_p=1$ and we get the theory
\begin{equation}
    \mathcal L_{1/3} = \frac{1}{4\pi}\text{Tr}\left[ \alpha d\alpha + \frac23 \alpha^3\right] + \frac{1}{4\pi}\text{Tr}\alpha d\text{Tr}\alpha+\frac{1}{2\pi}\text{Tr}\alpha~ dA,
\end{equation}
which we call $\U(2)_{-1,-6}=[\SU(2)_{-1}\times \U(1)_{-6}]/\mathbb Z_2$ where the quotient is taken by condensing the simple current boson labelled by $(1/2,3)$. The resulting theory has three anyons, $I,q,q^2$ with $q^3=I$ and where $q,q^2$ both have topological spin $\theta_q=e^{-2\pi i/3}$. The $Q=\frac13$ anyons in our theory are sourced by a combination of $q$ and the parton $f_{1,2}$ giving the full excitation spin $e^{2\pi i/6}$ as is familiar from $\U(1)_{3}$.

Now consider introducing a finite density of quasiholes, and assume we are in the intermediate case which leads to partons at filling $\tilde\nu=(-3,-3,0)$. The matter field associated with these holes so far has been $f\sim (f_1,f_2)$, but now we perform another partonization $f = \Phi \Psi$ where $\Phi$ is a boson and $\Psi$ a fermion. The idea is to first consider the theory with $\Phi$ gapped (the ``stacking'') and then condense $\Phi$, where we note that if $\Phi$ is condensed, $f$ and $\Psi$ are the same. Thus the final theory after this procedure will be exactly as discussed, and the introduction of $\Phi$ is just an analytical device to separate the stacking and condensation parts of the procedure.

The additional partonization introduces a new gauge field, $\beta$, which we take to be $\U(2)$ as well. We let $\Psi$ transform in the fundamental rep. of $\beta$ ($\Psi\rightarrow U_\beta \Psi)$, which means that $\Phi$ transforms under gauge transformations as $\Phi\rightarrow U_\alpha \Phi U_\beta^\dagger$, putting it in the fundamental of $\alpha$ and anti-fundamental of $\beta$ which ensures that $f\sim \Phi\Psi$ transforms as $f\rightarrow U_\alpha f$ as required. If $\Phi$ condenses, the Higgs mechanism necessitates that $\alpha=\beta$. Before proceeding, we need to choose a mean-field \textit{Ansatz}. To be able to condense $\Phi$ later on, we choose $\langle \alpha \rangle=\langle \beta \rangle$, meaning that $\Psi$ (the $\U(2)$ fermion) is at filling $\tilde\nu=-3$. Integrating $\Psi$ out (it is assumed gapped), we arrive at
\begin{equation}
    \mathcal L = \frac{1}{4\pi}\text{Tr}\left[ \alpha d\alpha + \frac23 \alpha^3\right] + \frac{1}{4\pi}\text{Tr}\alpha d\text{Tr}\alpha+\frac{1}{2\pi}\text{Tr}\alpha~ dA + \mathcal L[\Phi;\alpha,\beta] - \frac{3}{4\pi}\text{Tr}\left[ \beta d\beta + \frac23 \beta^3\right].
\end{equation}
Ignoring the matter field for now, this has stacked the original (charged) $\U(2)_{-1,-6}$ order with another (charge-neutral) topological order (described by $\beta$) which in this notation is $\U(2)_{3,6}$. Before proceeding with the condensation, we describe the anyon content of $\U(2)_{3,6}=[\SU(2)_3\times \U(1)_6]/\mathbb Z_2$. The anyons in $\SU(2)_{3}\times \U(1)_6$ are labelled as $(j,n)$ by the $\SU(2)$ spin $0\leq j\leq 3/2$ and $\U(1)$ charge $0\leq n<6$. There are no self-bosons in this theory, but $(3/2,3)$ is a fermion and the $\mathbb Z_2$ quotient describes stacking with the trivial fermionic theory $\{1,\tilde c\}$ and condensing $J=\tilde c\times (3/2,3)$ which is a bosonic simple current with $J^2=I$.

The resulting theory has 12 particle types \cite{jndb-435f} and is generated by $h,c,\tau$ where $h^3=I,~c^2=I$ and $\tau\times\tau=I+\tau$. In terms of $\U(2)$ representations $(j,n)$, we have $h=(0,2)$, $c=(3/2,3)$ and $\tau=(1,0)$ where the $\mathbb Z_2$ quotient necessitates $j+n/2\in\mathbb Z$. From these assignments, we see that $\Phi$, which transforms in the anti-fundamental representation $(1/2,5)$ of $\U(2)$, sources the anyon $v= hc\tau$ in this topological order. But note that $\Phi$ also transforms in the fundamental rep. of $\U(2)_{-1,-6}$ and as such sources the charge-$\frac13$ Laughlin anyon $\bar q$. When $\Phi$ condenses (by an appropriate choice of $\mathcal L[\Phi;\alpha,\beta]$), the physical picture is that upon doping an anyon $q$ into the system, this anyon gives its charge to the $\Phi$ condensate, which makes it transition into a neutral anyon in the stacked order, $q\rightarrow q(\bar q hc\tau)=hc\tau$. 

To continue the analysis, we consider the fusion rule $\tau\times\tau=I+\tau$. It has been argued in Ref.~\cite{senthil_doping_2025} that at a finite anyon density, the two channels cease to be degenerate. While the favoured channel depends on energetics, it appears more natural that the ``paramagnetic'' $I$ [carrying $j=0$ under $\SU(2)$] is favoured over the ``ferromagnetic'' $\tau$ (carrying $j=1$), with the argument coming from a calculation similar to that of Sect.~\ref{sup:su_m_csm}. Assuming that the $I$ is indeed favoured, we see that introducing three quasiholes leads to $v^3\sim c\tau$, which is a non-trivial anyon (the fermion $c$, together with the anyon $\tau$). To create a local excitation, we must introduce six quasiholes as $v^6\rightarrow I$. But such an excitation comprising of six quasiholes is a charge-$2e$ boson, the Cooper pair. 
If instead the $\tau$ channel was favoured, $v^3\sim c$ is a fermion, and this means that three quasiholes can lead to a local fermionic excitation (the electron). In that case, superconductivity might not be expected -- indeed, if the $\tau$ channel is favoured [due to the finite-$j$ representations of $\SU(2)$], the flux seen by $f_1,f_2$ would differ, which contradicts our numerical results in that regime, where equality of the mean-field fluxes, $b_1=b_2$, is favoured. Thu,s our numerical results are consistent with the above, $I$ case (and resulting Cooper pairs) and not this alternative $\tau$ scenario with low-energy fermions.

But because $\bar q hc\tau$ is not a bosonic simple current, it is not clear how to interpret the above argument in terms of anyon condensation. One would need anyon condensation to lead from $\U(2)_{-1,-6}\times U(2)_{3,6}$ to $\U(2)_{2,0}$ which has no residual anyons. A more detailed analysis would be required to understand the connection between the two.

Finally, we note the connection between our theory and the anyon superconductor of Ref.~\cite{seo_unified_2026}, where it is obtained by stacking $\U(1)_2\times \U(1)_{-6}$ on top of $\U(1)_{3}$ order. In our construction, start by considering the fully polarized case with filling fractions $\tilde\nu=(-\frac32,0,0)$. We have so far thought of this theory as a ``Secondary CFL'' following \cite{senthil_doping_2024}, but other gapped states are possible. Denoting the quasihole creation operator by $\psi$ (recall that it couples to a gauge field $a$), let us do another partonization $\psi=\Phi\varphi\Psi$ where $\Phi,\varphi$ are bosons coupling to $a-\alpha-\beta$ and $\alpha$ and $\Psi$ is a fermion coupling to $\beta$ with $\alpha,\beta$ being new gauge fields associated with this partonization. The idea is again to demand condensation of $\Phi$, meaning that we look for gapped mean fields where $\langle a\rangle=\langle \alpha+\beta\rangle$. One such choice is $\alpha=3a/4$ and $\beta=a/4$ which allows us to put $\varphi$ into the gapped $n=-2$ bosonic IQHE state (which is only trivially gapped if translation symmetry is broken) and $\Psi$ into the $n=-6$ fermionic IQHE. The resulting theory is 
\begin{equation} \label{eq:otherstack}
    \mathcal L = \frac{2}{4\pi}ada+\frac1{4\pi}adb+\frac{2}{4\pi}bdb + \mathcal L[\Phi,a-\alpha-\beta]-\frac{2}{4\pi}\alpha d\alpha - \frac6{4\pi}\beta d\beta,
\end{equation}
showing that we are stacking $\U(1)_2\times \U(1)_6$. Here, condensation of $\Phi$ has the clear interpretation of stacking the anyon $\bar q$ with $s$ (the semion in $\U(1)_2)$ and $\rho$ (from $\U(1)_{6}$ with $\alpha^6=I$). The composite $\bar q s\rho$ is a charged self-boson and may be condensed, which confines all remaining anyons and leads to a superconducting state. Note that here, introducing three $q$ quasiholes leads to $q^3(\bar q^ s\rho)^3=\rho^3\neq I$ which is an anyon, demonstrating that the lowest-charge local excitation consists of 6 quasiholes, i.e., it is the charge-$2e$ Cooper pair.

The slight difference between the descriptions arises because \cite{seo_unified_2026} begins with $\U(1)_3$ CS theory (with anyons sourced by a bosonic field), while our starting point is the $\U(1)^2$ theory with $K=\begin{pmatrix} 2 & 1\\1 & 2
\end{pmatrix}$, where the anyons are sourced by fermionic fields (the partons). To make the connection more explicit, one may perform a 3D XY duality replacing  $\Phi$ with a $\U(1)$ gauge field $c$ in Eq.~\ref{eq:otherstack} as $\mathcal L[\phi,a-\alpha-\beta]\rightarrow\frac{1}{2\pi} cd[a-\alpha-\beta]$.

\section{The effect of interactions} \label{sup:interactions_effect}

In this section, we expand on our argument in Sect.~\ref{sub:beta_int} on how particle interactions may discriminate between different candidate states. We start by focusing on the simplest example of two quasiholes in the $\nu=\frac12$ bosonic FCI where we partonize $\psi_b=f_1f_2$. In a generalized Landau Level, the wavefunctions
\begin{equation} \label{eq:two_local_holes}
    \chi^{uv}(\{z_i\})=\prod_i(z_i-u)(z_i-v)\prod_{i<j}(z_i-z_j)^2 e^{-\sum_i K(z_i,z_i^*)}
\end{equation}
represent an un-normalized, overcomplete basis of all two-quasihole states below the many-body gap, where $u,v\in\mathbb C$ represent the locations of the quasiholes in the complex plane. As a consequence, all the parton states we consider can be expressed as $\psi=\int d^2ud^2v~g(u,v)\chi^{uv}$. In this language, states with different mean-field fluxes $b_p$ considered in Sect.~\ref{sub:mfdope} correspond to different allowed forms of $g(u,v)$ [which can be related to a quasihole pseudowavefunction]. The implications of the different $g(u,v)$ on the kinetic energy terms have already been discussed in Sect.~\ref{sub:mfdope} and here, we discuss the effect different $g(u,v)$ have on the interaction terms.

In the quantum Hall problem without a kinetic term for the quasiholes, we expect the holes to localize and to have $g(u,v)\sim f(u-u_0)f(v-v_0)$, with $f(z)$ some function peaked around the origin which decays on a scale comparable to the magnetic length. For large $|u_0-v_0|$, the two quasiholes are always well separated and interactions between them do not play a significant role. The situation changes when we introduce a dispersion for the quasiholes in an FCI --- now, the kinetic term makes it favourable for the quasiholes to delocalize, and have a non-zero amplitude $g(u,v)$ for $u,v$ throughout the sample. Two such delocalized quasiholes will generally have an amplitude to be close to each other, and it is in this case that the interactions may introduce an energy penalty. To understand the energetics, we need to see how a delocalized $g(u,v)$ behaves when $u\rightarrow v$.

In the two-quasihole $\nu=\frac12$ state, there exist two possibilities for the parton flux -- either (A) both partons see equal flux, leading to one hole per parton, or (B) there is a re-distribution of flux, giving two holes to one parton and zero to the other. If we let $\ket{\text{LLL}_N}$ denote a filled LLL with $N$ orbitals, we can write down the mean-field parton states $\ket{\Omega_p}$ for the two cases as
\begin{align}
    \text{(A): }& \ket{\Omega_1}= f_\alpha\ket{\text{LLL}_N},~~\ket{\Omega_2}=f_\beta\ket{\text{LLL}_N}\\
    \text{(B): }& \ket{\Omega_1}= \tilde f_\alpha \tilde f_\beta\ket{\text{LLL}_{N+1}},~~\ket{\Omega_2}=\ket{\text{LLL}_{N-1}}
\end{align}
where the $f$'s are \textit{delocalized} parton annihilation operators. Note that all $\ket{\Omega_p}$ contain exactly $N-1$ particles. By writing $f_\alpha = \int d^2u f_\alpha(u) f(u)$ (with $f(u)$ the annihilation operator at position $u$), we get for example $\Omega_1^{(A)}=\int d^2u f_\alpha(u)\prod_{i}(u-z_i) \prod_{i<j}(z_i-z_j)$ and a similar expression for $\Omega_2^{(A)}$. The more tricky one is $\Omega_1^{(B)}$, where we write again by decomposing $\tilde f_{\alpha,\beta}$ in terms of local annihilation operators (with $\ket{z_1,\ldots,z_K}$ a state with localized partons)
\begin{equation}
\begin{split}
    \Omega_1^{(B)}(z_1,\dots,z_{N-1}) &= \braket{z_1z_2\dots z_{N-1}|\tilde f_\alpha \tilde f_\beta|\text{LLL}_{N+1}}=\int d^2u d^2v \tilde f_\alpha(u)\tilde f_\beta(v)\braket{z_1\ldots z_{N-1}uv|\text{LLL}_{N+1}}= \\
    &= \int d^2u d^2v \tilde f_\alpha(u)\tilde f_\beta(v) (u-v)\prod_{i}(u-z_i)(v-z_j) \prod_{i<j}(z_i-z_j).
\end{split}
\end{equation}
Using Eq.~\ref{eq:general_parton_to_wf}, we find that these forms imply the microscopic wavefunctions (written in terms of Eq.~\ref{eq:two_local_holes} and using the symmetry $\chi^{uv}=\chi^{vu}$)
\begin{align}
    \text{(A): }& \psi({z_1},\ldots,z_{N-1})=\mathcal N ^\text{(A)}\int d^2ud^2v \left[f_\alpha(u)f_\beta(v)+f_\alpha(v)f_\beta(u)\right] \chi^{uv}(z_1,\ldots,z_{N-1}),\label{eq:fullwfA}\\
    \text{(B): }& \psi({z_1},\ldots,z_{N-1})=\mathcal N ^\text{(B)}\int d^2ud^2v \left[\tilde f_\alpha(u)\tilde f_\beta(v)-\tilde f_\alpha(v)\tilde f_\beta(u)\right]~[u-v]~\chi^{uv}(z_1,\ldots,z_{N-1})\label{eq:fullwfB}
\end{align}
with $\mathcal{N}^{\text{(A),(B)}}$ normalization constants. From this, the two forms for $g(u,v)$ can be read off as $g^{\text{(A)}}(u,v)=\mathcal N ^\text{(A)}[f_\alpha(u)f_\beta(v)+f_\alpha(v)f_\beta(u)]$ and $g^\text{(B)}(u,v)=\mathcal N ^\text{(B)}(u-v)[\tilde f_\alpha (u)\tilde f_\beta(v)-\tilde f_\alpha (v)\tilde f_\beta(u)]$. We now make the crucial assumption of quasihole delocalization -- that is, for a generic position $z$, all the orbitals $f_{\alpha,\beta}(z),~\tilde f_{\alpha,\beta}(z)$ have non-zero magnitude. In that case, we see that $g^\text{(A)}(u,v)\sim \text{const.}$ as $u\rightarrow v$, demonstrating that $\chi^{u\rightarrow v,v}$ is represented in Eq.~\ref{eq:fullwfA} as well as states with the holes well-separated. On the other hand, we can clearly see that $g^\text{(B)}(u,v)\rightarrow 0$ when $u\rightarrow v$, even if $\tilde f_{\alpha,\beta}$ are both extended. This shows that $\chi^{u\rightarrow v,v}$ is under-represented in Eq.~\ref{eq:fullwfB}, relative to states with the holes further separated, implying a suppression for the amplitude of the quasiholes approaching each other. Assuming that a repulsive interaction between the microscopic particles leads to a repulsive effective interaction between the quasiholes, we then see that such interactions would favour the polarized scenario (B) because it prevents the quasiholes from approaching each other, unlike (A).

We note that the delocalization of quasiholes is necessary for this argument to work. If on the contrary $f_{\alpha,\beta},\tilde f_{\alpha,\beta}$ were localized around $w_\alpha,w_\beta$, the two forms would be closely related by the choice of normalization $\mathcal N^\text{(A)}=(w_\alpha-w_\beta)\mathcal N^\text{(B)}$. When the $f$'s are delocalized, however, the forms of $g$ are truly different as described above. Thus, delocalisation of the quasiholes is crucial for this interaction effect to be significant, and we thus do not expect to see any analogous effect in pure quantum Hall physics.

A comment on the exchange symmetry of the quasiholes is in order. When adiabatically exchanging two quasiholes in the $\nu=\frac12$ FCI, we expect to obtain the semionic phase $e^{i\pi/2}$, but this phase is not respected in our construction. To rectify this, we first stress that the \textit{Ansätze} Eq.~\ref{eq:fullwfA},~\ref{eq:fullwfB} crucially respect the correct (bosonic) exchange statistics for the actual microscopic particles, meaning they are valid as variational wavefunction forms and their quality should be judged as a function of their energy. But more physically, the absence of this phase may be tied to the charge-flux unbinding of the quasiholes \cite{senthil_doping_2025}. A true semion would correspond to locally acting with $f_p(\mathbf r_0)$ to remove a parton \textit{and} introducing a $\pi$-flux of the internal gauge field at the point $\mathbf r_0$. These fluxes associated with the quasiholes are responsible for the semionic exchange statistics. But in our construction, we imagine a uniform flux of the internal field. The idea is that at a finite density of quasiholes, we may replace the fluctuating fluxes tied to the quasiholes by an average background mean-field, in an approximation exactly analogous to the usual ``flux smearing'' done in the composite fermion approach to quantum Hall states. Because of this smearing, which is crucial to get anyon-superconducting behaviour, the quasiholes considered in Eq.~\ref{eq:fullwfA},~\ref{eq:fullwfB} are not exactly semions (which would be the bound state of a parton hole and a $\pi$-flux), but are instead only the ``charge'' part of this charge-flux composite and as such, they are not expected to obey semionic exchange statistics.  Due to this unbinding, the resulting superconductor no longer supports anyons as excitations.

The crucial ingredient in this calculation was the fact that the fermionic nature of the parton holes suppresses the amplitude for two such holes to approach each other when the holes arise from the same parton (B). When arising from different partons (A), there exists no such suppression. This principle trivially generalizes to many quasiholes and more species of partons, as explained in Sect.~\ref{sub:beta_int}.

\section{Smaller doping fraction for the fermionic $\nu=\frac13$ FCI} \label{sub:smallerdope}
\begin{figure}
    \centering
    \includegraphics[width=0.99\linewidth]{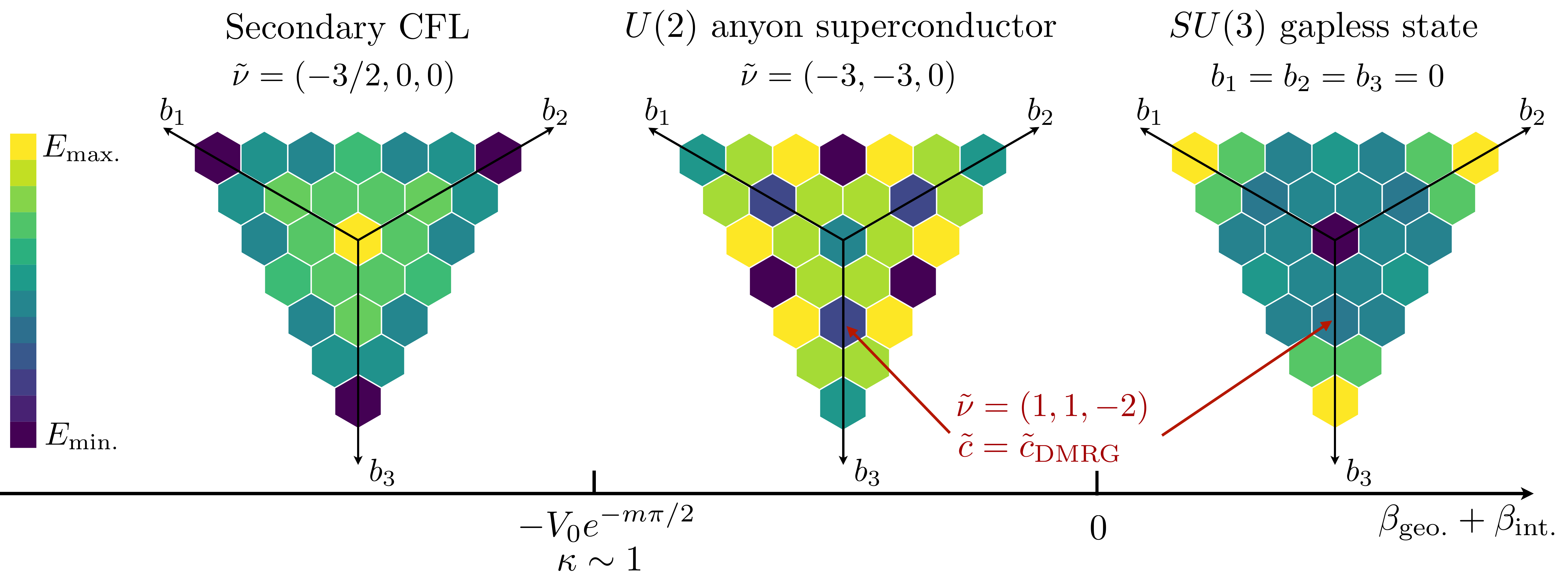}
    \caption{States of the doped $\nu=\frac13-\delta$ fermionic FCI at $\delta=\frac1{18}$ on a $6\times6$ torus for parameters otherwise identical to Fig.~\ref{fig:fermions_results}. The results and different anyon phases here are consistent with the picture observed at $\delta=\frac19$.}
    \label{fig:energy_fermi_smalldope}
\end{figure}

In the main text, we considered $\nu=\frac13-\delta$ with $\delta=1/9$. In this section, an analogous calculation is performed for $\delta=1/18$ with the result shown in Fig.~\ref{fig:energy_fermi_smalldope}. The results are consistent with what was discussed in the main text. The model and parameters used are identical to Fig.~\ref{fig:fermions_results} in the main text. Note that there are fewer possible values of $b_p$ here, as the variation of $b_p$ cannot be arbitrarily fine --- it is set by the magnetic monopole quantization constraint that $2\pi b_p N_s\in\mathbb Z$ for a torus with $N_s$ sites.

\section{Non-Abelian Chern-Simons-Maxwell theory} \label{sup:su_m_csm}
Our VMC numerical results indicate that there exist systems where the best parton description of the doped state has $b_1=\ldots=b_m=0$ and the mean-field \textit{Ansatz} preserves the full $\SU(m)$ gauge invariance. The partons in these states are gapless, and to understand what these states physically correspond to, interactions between partons, mediated by the gauge field, become important. 

The system may be described as as a small density $m\delta$ of quasiholes $f^\dagger =(f_1^\dagger,f_2^\dagger,\ldots f_m^\dagger)$ transforming in the fundamental representation of an $\SU(m)$ gauge field $a_\mu$. Following \cite{senthil_doping_2025}, the effective theory contains (1) level-1 self-Chern Simons term for $a_\mu$ as a result of the Laughlin state we are doping on top of, (2) Some action $\mathcal L[f,a]$ for the quasiholes coupled to the gauge field and (3) less RG-relevant terms, including a Maxwell-Type term for $a_\mu$,
\begin{equation}
    \mathcal L = -\frac{1}{4\pi}\text{Tr}\left(ada+\frac23 a^3\right)+\frac{1}{2g^2}\text{Tr} \mathfrak{f}\wedge \star \mathfrak{f} + \mathcal L[f,a]
\end{equation}
with $\mathfrak f=da+a^2$. Integrating out the gauge field $a$ will lead to an effective interaction term for $f$. The combination of the Chern-Simons and Maxwell terms endows $a$ with a mass $\mu=g^2/2\pi$ and the resulting interaction potential will decay with separation. The more subtle question is the sign of the interaction. To understand this, let  $\mathcal T^{a}$ ($a=1,\ldots, m^2-1$) be the Lie algebra generators of $\SU(m)$. The effective two-particle interaction looks like \cite{senthil_doping_2025}
\begin{equation}
    U(\mathbf r) \sim \frac{e^{-\mu|\mathbf r|}}{\sqrt{|\mathbf r|}} \sum_a \mathcal T^a_{(1)}\otimes \mathcal T^a_{(2)}
\end{equation}
where the $\mathcal T_{{(1),(2)}}$ act on the two particles (each of $T^a$ is an $m\times m$ matrix, and a Kronecker product between the two is understood). The negative eigenvalues of the matrix on the right side (and their eigenvectors) will determine in what states two quasiholes must be to be attracted by this potential. 

Consider first $\SU(2)$. Combining two fundamental (spin-$\frac12$) particles gives 4 eigenvalues, the singlet at $-3$ and the triplet at $+1$. This means that the particles are only attracted when in the singlet channel, and singlet bound states may form. For $m=2$, this physically corresponds to pairing $\sim \langle f_1f_2\rangle$ as discussed in the main text. The resulting bound state is neutral with respect to the $\SU(2)$ field and will be invisible to newly added quasiholes.

At the level of representation theory, we have $\mathbf 2\times\mathbf 2=\mathbf1+\mathbf3$ and we are evaluating the quadratic Casimir for the irreps on the right side. The same game may be played for $\SU(3)$, where $\mathbf 3\times\mathbf 3=\bar{\mathbf 3}+\mathbf 6$ with the anti-fundamental $\bar{\mathbf 3}$ having lower energy, although it is still charged. Upon adding a third particle, $\mathbf{3}\times\mathbf{3}\times\mathbf{3}=\mathbf1+\mathbf8+\mathbf8+\mathbf{10}$ with the singlet $\mathbf 1$ having the lowest value of $\sum_a\left[\mathcal T^a_{(1)}\mathcal T^a_{(2)}I_{(3)}+\mathcal T^a_{(1)}I_{(2)}\mathcal T^a_{(3)}+I_{(1)}\mathcal T^a_{(2)}\mathcal T^a_{(3)}\right]$ (Kronecker products implied), meaning the strongest attractive interaction potential $U(\mathbf r)$. Again, groups of three quasiholes may be brought together to form a bound singlet state which is then neutral with respect to $a_\mu$, meaning that we should not expect the energy to be further lowered by interactions with more particles. 

We stress that this is merely a tree-level perturbative analysis of the inter-parton interactions mediated by the gauge fields. The many-body parton state will be determined by the competition between interactions of multiple partons and also the competition between kinetic energy and these interactions. Informed by the nature of the interaction, we suggested some possible conclusions in the main text, but a final answer to this question is beyond the scope of this work.